\documentclass{article}

\usepackage[final]{neurips_2024}

\usepackage[utf8]{inputenc} 
\usepackage[T1]{fontenc}    

\usepackage{url}            
\usepackage{booktabs}       
\usepackage{amsfonts}       
\usepackage{nicefrac}       
\usepackage{microtype}      
\usepackage{xcolor}         
\usepackage{graphicx}
\usepackage[textsize=footnotesize]{todonotes}
\usepackage{xcolor}
\usepackage{amsmath}
\usepackage{listings}
\usepackage{spverbatim}
\usepackage{hyperref} 
\hypersetup{
    colorlinks=true,    
    linkcolor=blue,     
    citecolor=blue,    
    urlcolor=blue
}
\usepackage[nameinlink]{cleveref}
\usepackage{tikz}
\usepackage{forest}
\usepackage{subcaption}
\usepackage{dirtree}
\usepackage{mdframed}
\usepackage{tabularx}

\presetkeys{todonotes}{disable}{}

\usepackage{natbib}

\crefname{table}{Table}{Tables}
\crefname{section}{Section}{Sections}
\crefname{figure}{Figure}{Figures}

\title{Which Economic Tasks are Performed with AI? \\Evidence from Millions of Claude Conversations}

\author{
  Kunal Handa\thanks{Equal contribution. Authors above the line break are core contributors. Author contributions are listed in \Cref{appendix:contributions}. Direct correspondence to \{kunal, atamkin, deep\} @anthropic.com.},\, Alex Tamkin\footnotemark[1],\, Miles McCain, Saffron Huang, Esin Durmus\\\\
  \textbf{Sarah Heck, Jared Mueller, Jerry Hong, Stuart Ritchie, Tim Belonax, Kevin K. Troy}\\\\
  \textbf{Dario Amodei, Jared Kaplan, Jack Clark, Deep Ganguli}\\\\
  Anthropic\\
}

\makeatletter
\def\@noticestring{}
\makeatother

\begin{document}
\maketitle

\begin{abstract}
Despite widespread speculation about artificial intelligence's impact on the future of work, we lack systematic empirical evidence about how these systems are actually being used for different tasks. Here, we present a novel framework for measuring  AI usage patterns across the economy. We leverage a recent privacy-preserving system \citep{clio} to analyze over four million Claude.ai conversations through the lens of tasks and occupations in the U.S. Department of Labor's O*NET Database. Our analysis reveals that AI usage primarily concentrates in software development and writing tasks, which together account for nearly half of all total usage. However, usage of AI extends more broadly across the economy, with $\sim\!36\%$ of occupations using AI for at least a quarter of their associated tasks. We also analyze \textit{how} AI is being used for tasks, finding 57\% of usage suggests augmentation of human capabilities (e.g., learning or iterating on an output) while 43\% suggests automation (e.g., fulfilling a request with minimal human involvement). While our data and methods face important limitations and only paint a picture of AI usage on a single platform, they provide an automated, granular approach for tracking AI's evolving role in the economy and identifying leading indicators of future impact as these technologies continue to advance.

\end{abstract}

\begin{figure}
    \centering \includegraphics[width=0.99\linewidth]{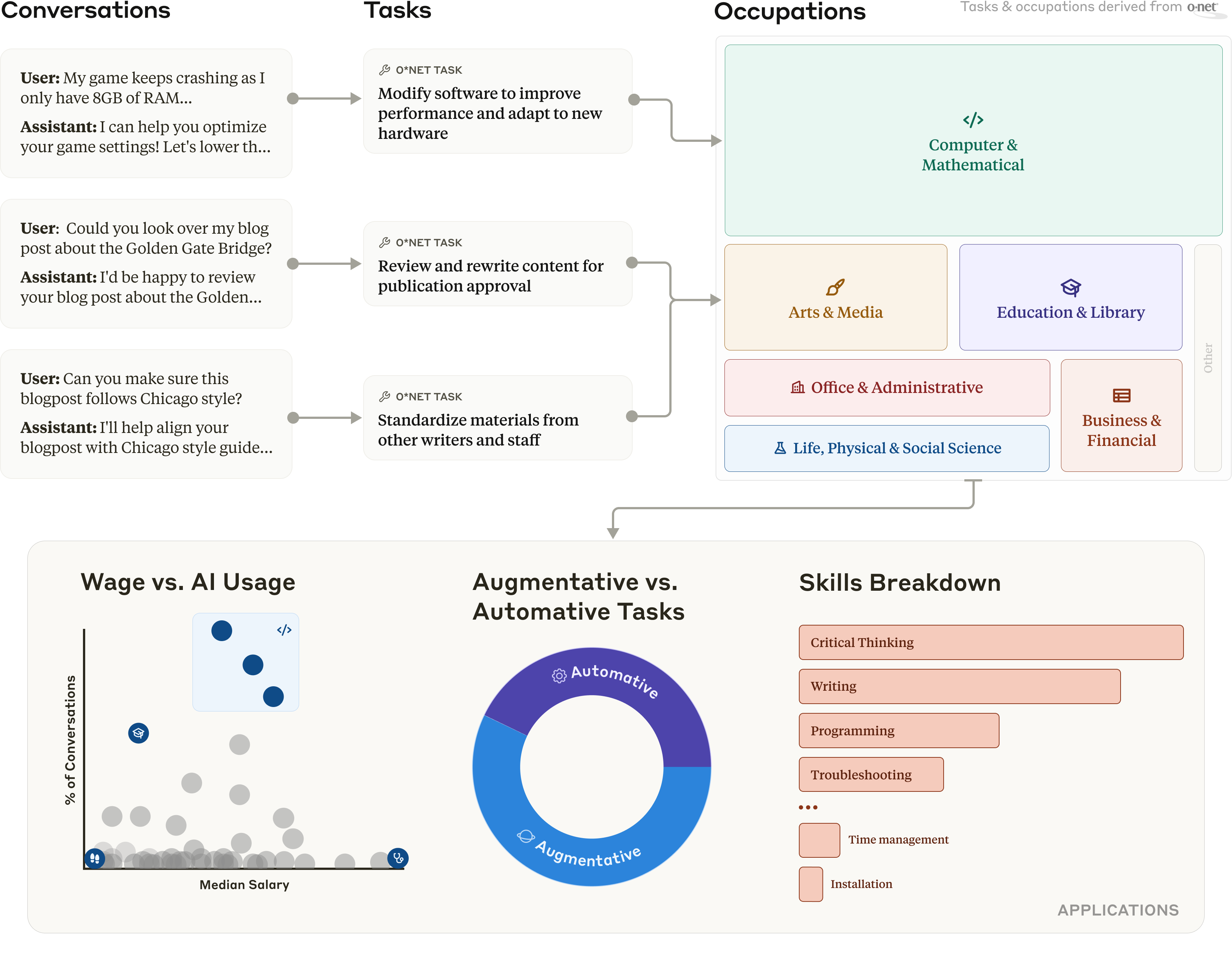}
    \caption{\textbf{Measuring AI use across the economy.} We introduce a framework to measure the amount of AI usage for tasks across the economy . We map conversations from Claude.ai to occupational categories in the U.S. Department of Labor's O*NET Database to surface current usage patterns. Our approach provides an automated, granular, and empirically grounded methodology for tracking AI's evolving role in the economy. \textit{(Note: figure contains illustrative conversation examples only.)}}
    \label{fig:pull}
\end{figure}

\section{Introduction}
Rapid advances in artificial intelligence suggest profound implications for the evolution of labor markets \citep{
brynsml, acemoglu2021harms, KorniekGrowth, indeedAI, comunale2024economic, maslej2024artificialintelligenceindexreport}. 
Despite the importance of anticipating and preparing for these changes, we lack systematic empirical evidence about how AI systems are actually being integrated into the economy. Existing methodologies---whether developing predictive models \citep{Webb2019TheIO, gptsaregpts, kinder2024generative}, conducting controlled studies of productivity effects \citep{peng2023impact, noy2023experimental}, or administering periodic surveys of users \citep{humlum2024adoption,bick2024rapid}---cannot track the dynamic relationship between advancing AI capabilities and their direct, real-world use across the economy.
 
Here, we present a novel empirical framework for measuring AI usage across different tasks in the economy, drawing on privacy-preserving analysis of millions of real-world conversations on Claude.ai \citep{clio}. By mapping these conversations to occupational categories in the U.S. Department of Labor's O*NET Database, we can identify not just current usage patterns, but also early indicators of which parts of the economy may be most affected as these technologies continue to advance.\footnote{We provide relevant data at  \href{https://huggingface.co/datasets/Anthropic/EconomicIndex/}{https://huggingface.co/datasets/Anthropic/EconomicIndex/}.}

\begin{figure*}
    \centering
    \includegraphics[width=0.99\textwidth]{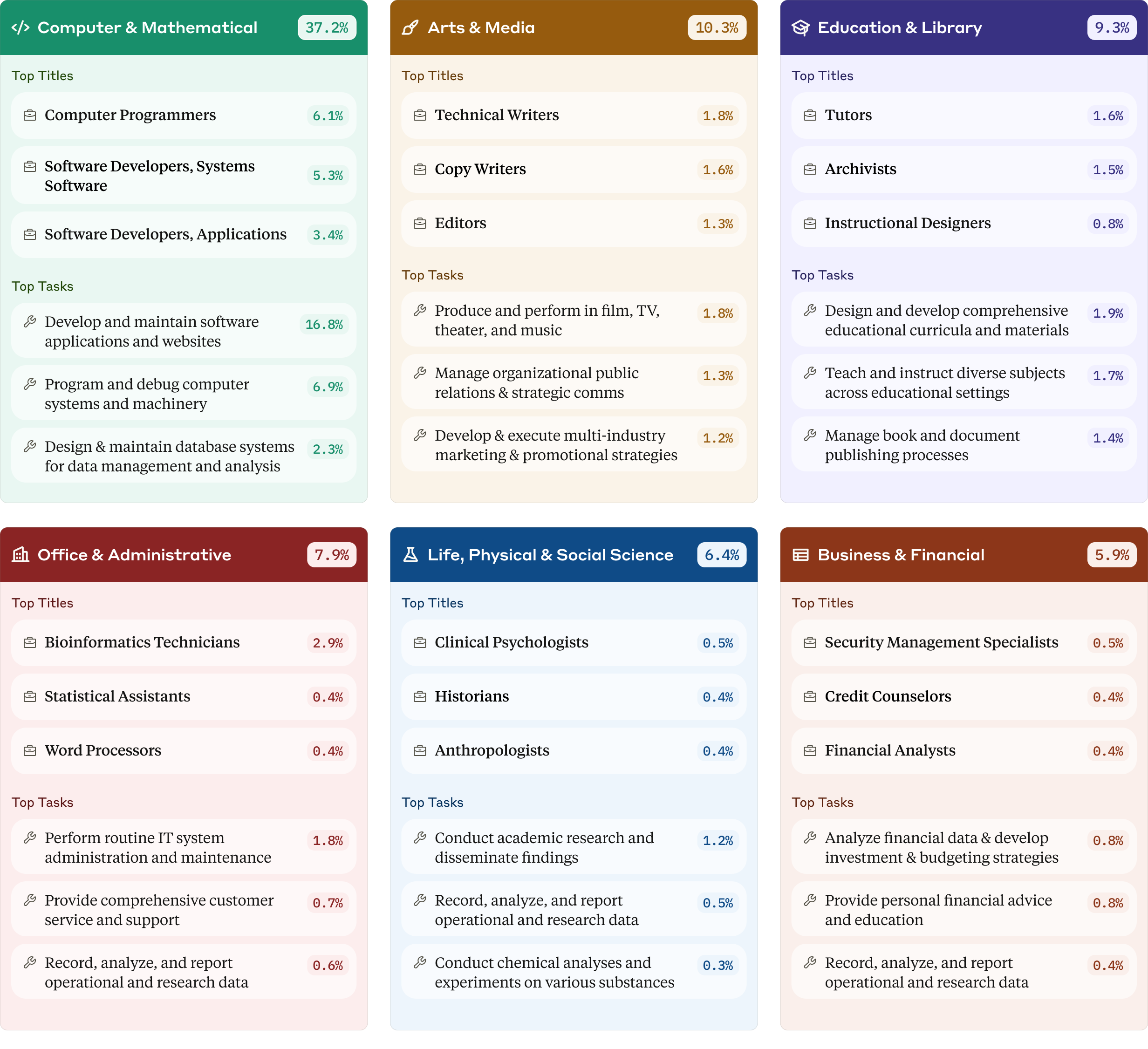}
    \caption{\textbf{Hierarchical breakdown of top six occupational categories by the amount of AI usage in their associated tasks.} Each occupational category contains the individual O*NET occupations and tasks with the highest levels of appearance in Claude.ai interactions.}
    \label{fig:occupations_and_tasks}
\end{figure*}

\begin{figure*}
    \centering
    \includegraphics[width=0.99\textwidth]{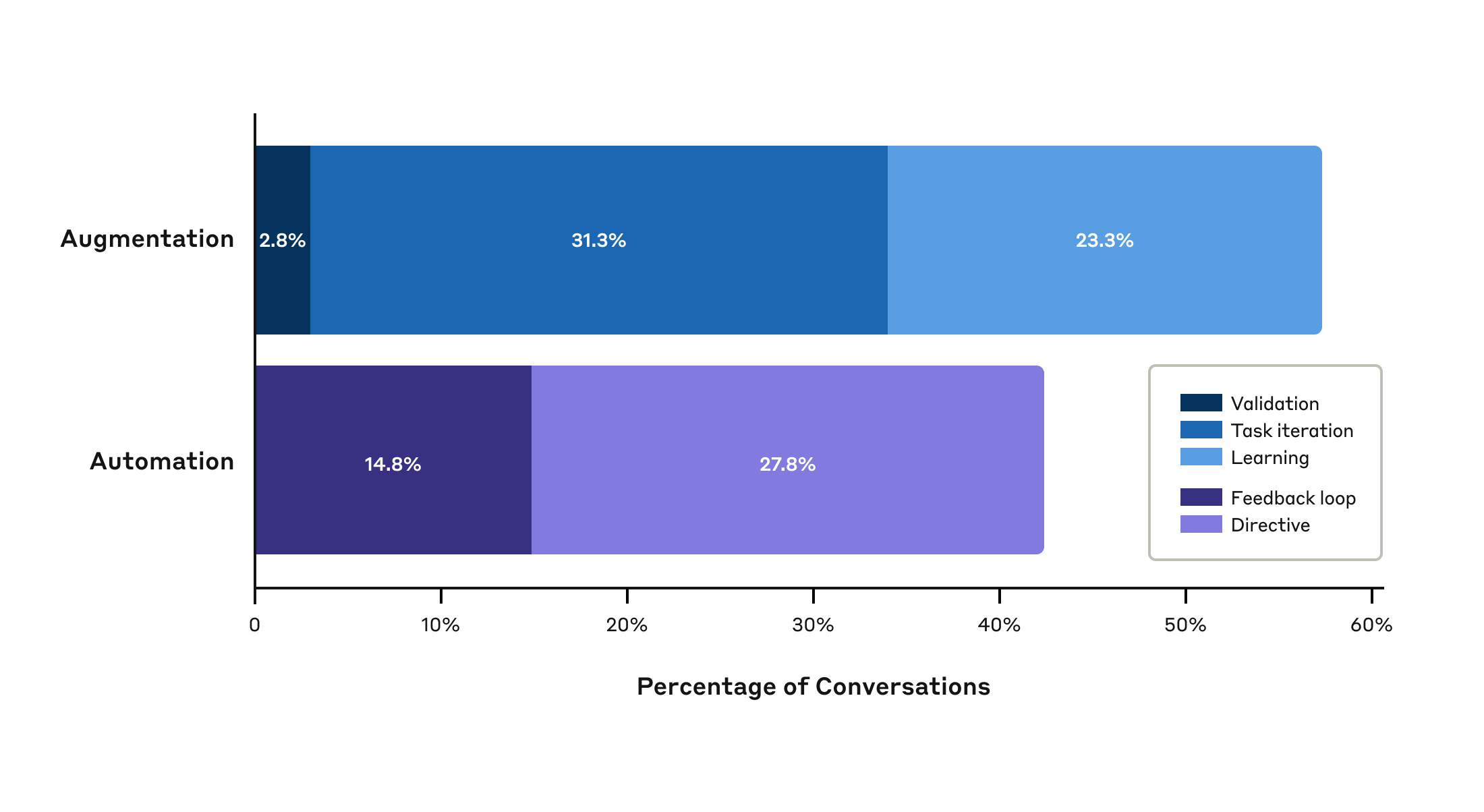}
    \caption{\textbf{Distribution of automative behaviors (43\%) where users delegate tasks to AI, and augmentative behaviors (57\%) where users actively collaborate with AI.} Patterns are categorized into five modes of engagement; automative modes include Directive and Feedback Loop, while augmentative modes are comprised of Task Iteration, Learning, and Validation.}
    \label{fig:autaug}
\end{figure*}

We use this framework to make five key contributions: 
\begin{enumerate}
    \item \textbf{Provide the first large-scale empirical measurement of which tasks are seeing AI use across the economy (\Cref{fig:pull}, \Cref{fig:occupations_and_tasks}, and \Cref{fig:occupations_usage})} Our analysis reveals highest use for tasks in software engineering roles (e.g., software engineers, data scientists, bioinformatics technicians), professions requiring substantial writing capabilities (e.g., technical writers, copywriters, archivists), and analytical roles (e.g., data scientists). Conversely, tasks in occupations involving physical manipulation of the environment (e.g., anesthesiologists, construction workers) currently show minimal use.
    \item \textbf{Quantify the depth of AI use within occupations (\Cref{fig:fraction_tasks_adopted})} Only $\sim\!4\%$ of occupations exhibit AI usage for at least 75\% of their tasks, suggesting the potential for deep task-level use in some roles. More broadly, $\sim\!36\%$ of occupations show usage in at least 25\% of their tasks, indicating that AI has already begun to diffuse into task portfolios across a substantial portion of the workforce.
    \item \textbf{Measure which occupational skills are most represented in human-AI conversations (\Cref{fig:skills_usage})}. Cognitive skills like Reading Comprehension, Writing, and Critical Thinking show high presence, while physical skills (e.g., Installation, Equipment Maintenance) and managerial skills (e.g., Negotiation) show minimal presence---reflecting clear patterns of human complementarity with current AI capabilities.
    \item \textbf{Analyze how wage and barrier to entry correlates with AI usage (\Cref{fig:wage_by_usage,tab:barrier_to_entry})}. We find that AI use peaks in the upper quartile of wages but drops off at both extremes of the wage spectrum. Most high-usage occupations clustered in the upper quartile correspond predominantly to software industry positions, while both very high-wage occupations (e.g., physicians) and low-wage positions (e.g., restaurant workers) demonstrate relatively low usage. This pattern likely reflects either limitations in current AI capabilities, the inherent physical manipulation requirements of these roles, or both. Similar patterns emerge for barriers to entry, with peak usage in occupations requiring considerable preparation (e.g., bachelor's degree) rather than minimal or extensive training.
    \item \textbf{Assess whether people use Claude to automate or augment tasks (\Cref{fig:autaug})} We find that 57\% of interactions show augmentative patterns (e.g., back-and-forth iteration on a task) while 43\% demonstrate automation-focused usage (e.g., performing the task directly). While this ratio varies across occupations, most occupations exhibited a mix of automation and augmentation across tasks, suggesting AI serves as both an efficiency tool and collaborative partner.
    
\end{enumerate}

Our methods provide an automated, granular, and empirically grounded approach for tracking AI usage patterns as both capabilities and societal usage evolve. This early visibility into emerging trends gives policymakers and civil society crucial lead time to respond to shifts in how AI transforms work. However, we acknowledge multiple key limitations (discussed in \Cref{sec:limitations}); for example, our usage data cannot reveal how Claude's outputs are actually used in practice, and our reliance on O*NET's static occupational descriptions means we cannot account for entirely new tasks or jobs that AI might create.\footnote{Though our methodology using Clio \citep{clio} will allow us to detect such emerging patterns of work as they arise.}

Nevertheless, this framework offers a foundation for understanding AI's evolving impact on the economy. While our methods are imperfect, they provide a systematic way to track usage patterns and identify leading indicators of economic effects across different sectors. 
As AI capabilities and adoption continue to advance, we believe this kind of empirical measurement will be crucial for understanding and preparing for the technology's broader economic implications.

\section{Background and Related Work}

Our work builds on many lines of research attempting to model, measure, and forecast AI's impact on the economy.

\paragraph{Economic foundations and the task-based framework}
A wide body of work in economics has proposed theoretical models to understand the impact of automation on the labor market. Most notably, \citet{autor2003skill, autor2013task} argue for modeling labor markets through the lens of discrete \textit{tasks} which can be performed by either human workers or machines---for example, \textit{debugging code} or \textit{cutting hair}. Building on this framework, \citet{autor2015there} shows that while technologies automate some tasks, they often augment human capabilities in others due to complementarity between humans and machines, leading to higher demand for labor. In addition \citet{acemoglu2018race} use this framework to explore a model where automation technologies can create entirely new tasks in addition to displacing old tasks. 

\paragraph{Forecasting the impact of AI on labor markets}
Another branch of work leverages the task-based framework to predict the future prevalence of automation across the economy, often based on descriptions of tasks and occupations from the O*NET database of occupational information provided by the U.S. Department of Labor \citep{onet2025}. For example, \citet{frey2017future} fit a gaussian process classifier to a dataset of 70 labeled occupations to predict which occupations are subject to computerization. \citet{brynjolfsson2018can} hire human annotators to rate 2,069 detailed work areas in the O*NET database, focusing specifically on their potential to be performed by machine learning. \citet{Webb2019TheIO} analyzes the overlap between patent documents and job task descriptions to predict the "exposure" of tasks to AI, finding highest exposure in high-education, high-wage occupations---a pattern partially reflected in our empirical usage data, though we find peak usage in mid-to-high wage occupations rather than at the highest wage levels. \citet{felten2023will} focus specifically on large language models, estimating exposure by using a dataset that links human abilities to different occupations. 

\citet{gptsaregpts} also consider exposure of tasks to language models, using language models themselves to obtain more granular estimates of exposure at the level of individual tasks---an approach we follow in our work. When considering the impacts of language model-powered software, they conclude that around half of all tasks in the economy could one day be automated by language models. While our empirical usage data shows lower current adoption ($\sim\!36\%$ of occupations using AI for at least a quarter of their tasks), the patterns of usage across tasks largely align with their predictions, particularly in showing high usage for software development and content creation tasks.

\paragraph{Real-world studies of AI usage}

To complement these forecasts based on human or machine judgment, another body of work attempts to gather concrete data to understand how AI is currently being adopted across the labor market. For example, studies show rapid AI adoption across different sectors and countries: research from late 2023 found that half of workers in exposed Danish occupations had used ChatGPT, estimating it could halve working times in about a third of their tasks \citep{humlum2024adoption}, while a subsequent study in August 2024 found that 39\% of working-age US adults had used generative AI, with about a quarter using it weekly \citep{bick2024rapid}. Moreover, further research has attempted to measure the breadth and depth of this usage, with studies finding positive effects of generative AI tools on productivity for a wide range of individual domains, including software engineering \citep{peng2023impact,cui2024effects}, writing  \citep{noy2023experimental}, customer service \citep{brynjolfsson2023generative}, consulting \citep[for tasks that were suitable for AI]{dell2023navigating}, translation \citep{merali2024scaling}, legal analysis \citep{choi2023ai}, and data science \citep{wiles2024genai}.

We bridge these separate approaches to perform the first large-scale analysis of how advanced AI systems are actually being used across tasks and occupations. We build on the task-based framework, but rather than forecasting potential impacts ("exposure" of occupations to AI), we measure real-world \textbf{usage patterns} using Clio \citep{clio}, a recent system that enables privacy-perserving analysis of millions of human-AI conversations on a major model provider. This allows us to complement controlled studies of AI productivity effects in specific domains with a comprehensive view of where and how AI is being integrated into work across the economy. Our methodology enables tracking these patterns dynamically as both AI capabilities and societal adoption evolve---revealing both present day usage trends as well as leading indicators of future diffusion.

\section{Methods and analysis} 

To understand how AI systems are being used for different economic tasks, we leverage Clio \citep{clio}, an analysis tool that uses Claude \citep{claudeCard} to provide aggregated insights from millions of human-model conversations. We use Clio to classify conversations across occupational tasks, skills, and interaction patterns, revealing breakdowns across these different categories. All analyses draw from conversation data collected during December 2024 and January 2025. See \Cref{appendix:experimental_details,appendix:prompts,appendix:metadata} for more details and prompts, including validating the composition of our dataset and how we perform classification in cases with large numbers of categories (e.g. O*NET tasks).

\begin{figure*}
    \centering
    \includegraphics[width=0.99\textwidth]{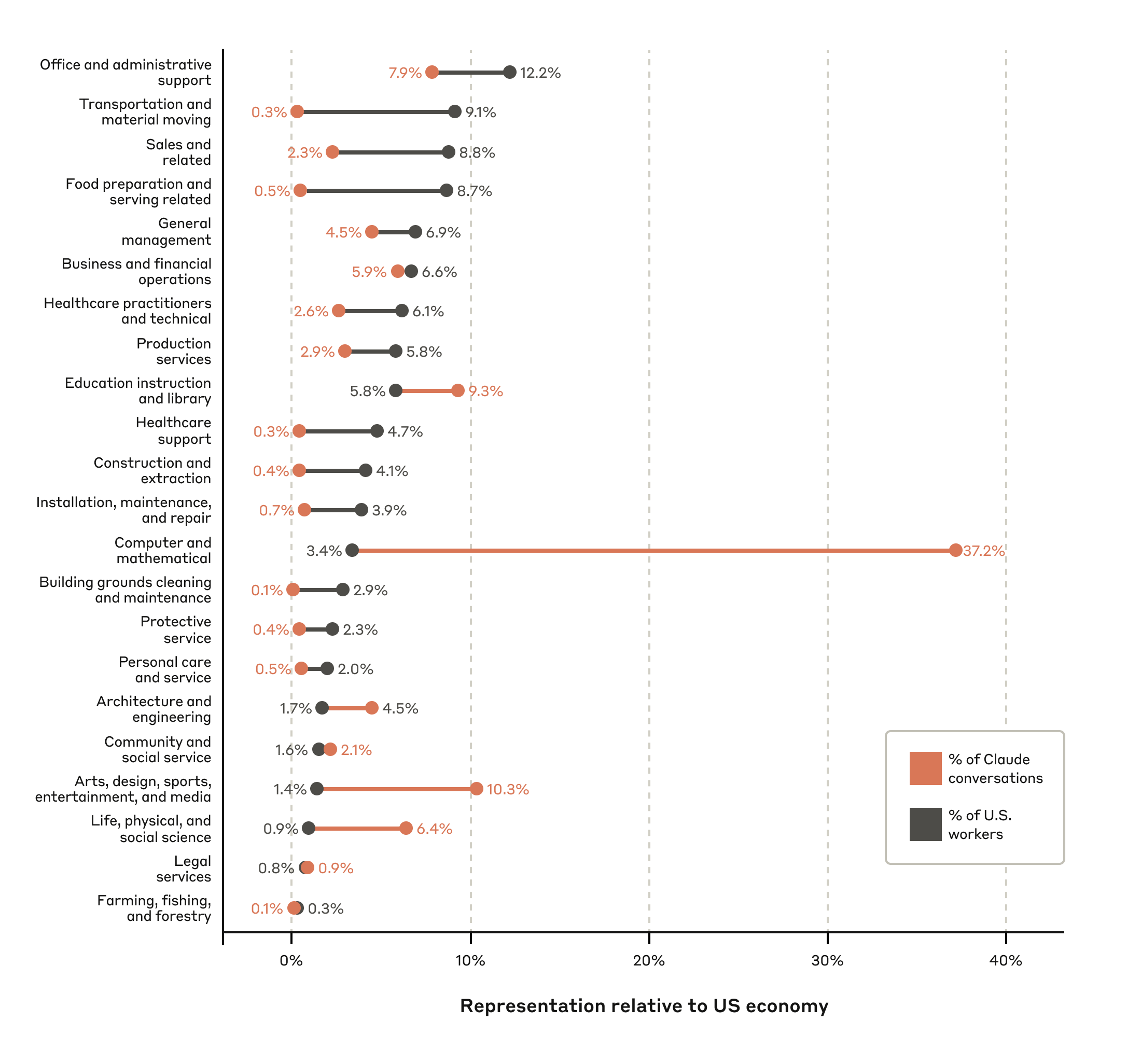}
    \caption{\textbf{Comparison of occupational representation in Claude.ai usage data and the U.S. economy.} Results show most usage in tasks associated with software development, technical writing, and analytical, with notably lower usage in tasks associated with occupations requiring physical manipulation or extensive specialized training. U.S. representation is computed by the fraction of workers in each high-level category according to the U.S. Bureau of Labor Statistics \citep{bls_website}.}
    \label{fig:occupations_usage}
\end{figure*}

\subsection{Task-level analysis of AI usage}
\label{sec:task_and_occupation_usage}

Using Clio on a dataset of one million Claude.ai Free and Pro conversations,\footnote{Anthropic enforces strict internal privacy controls. Our \href{https://www.anthropic.com/legal/privacy}{privacy policy} enables us to analyze \textit{aggregated and anonymized} user conversations to understand patterns or trends. We continue to manage data according to our \href{https://privacy.anthropic.com/en/articles/10023548-how-long-do-you-store-personal-data}{privacy and retention policies}, and maintain our approach of not training our generative models on user conversations by default. Because we focus on studying patterns in individual usage, the results shared in this paper exclude activity from business customers (i.e., Team, Enterprise, and all API customers). For more information, see Appendix F in \citet{clio}.} we analyzed each interaction to map it to its most relevant task category in the O*NET database. Because there are nearly 20,000 unique task statements in O*NET, we create a hierarchical tree of tasks using Clio, and perform the assignment by traversing the tree. Although there are often multiple valid tasks that a single conversation could be mapped to, we observed qualitatively very similar results when mapping a single conversation to multiple tasks.  We provide additional details and analyses in \Cref{appendix:experimental_details}, including for how we map conversations to tasks (\Cref{appendix:adoption_across_tasks}), the hierarchy creation process (\Cref{appendix:task-hierarchy}), the fact that we obtain similar results for conversation-level and account-level data (\Cref{appendix:conversations-vs-accounts}), and validation of our dataset composition (\Cref{appendix:dataset-composition}). Additionally, we discuss human validation of our task hierarchy classifications in \Cref{appendix:validation} and results on cluster-level data in \Cref{appendix:cluster_experiments}.

Our analysis reveals that computer-related tasks see the largest amount of AI usage, followed by writing tasks in educational and communication contexts. To understand broader patterns, we group these tasks according to O*NET's occupational framework---first mapping them to specific occupations (such as Computer Network Architects) and then to broader occupational categories (like Computer and Mathematical Occupations). \Cref{fig:occupations_and_tasks} illustrates this distribution across occupational categories, showing the most common occupations and tasks within each group, while \Cref{fig:occupations_usage} compares these usage patterns to the actual distribution of these occupations in the U.S. workforce.

It is worth noting that the occupational classification of a conversation does not necessarily mean the user was a professional in that field. For instance, while some nutrition-related queries might come from dietitians, others likely come from individuals seeking personal dietary advice. This widespread access to AI assistance for traditionally professional tasks---even if the assistance is not perfect---could have significant implications for these fields, though analyzing such impacts lies outside the scope of this study. We discuss these limitations further in \Cref{sec:limitations}.

Overall, the data reveal several insights:
\begin{itemize}
    \item In line with our task-level findings, Computer and Mathematical occupations show the highest associated AI usage rate, comprising 37.2\% of all queries. 
    \item Arts, Design, Entertainment, Sports, and Media occupations were second most common (10.3\%), likely reflecting the prevalence of marketing, writing, and other kinds of content generation in Claude.ai traffic. 
    \item Education occupations were also highly represented, including under Educational Instruction and Library Occupations as well as subject-specific occupations such as Life, Physical, and Social Science Occupations. 
    \item A significant portion of Claude.ai traffic fell under business-relevant occupations: Business and Financial Occupations, Office and Administrative Support Occupations, and Management Occupations. 
    \item Unsurpringly, occupations requiring physical labor were least present, for example Transportation and Material Moving Occupations; Healthcare Support Occupations; and Farming, Fishing, and Forestry Occupations. 
\end{itemize}

\begin{figure}
    \centering
    \includegraphics[width=0.9\linewidth]{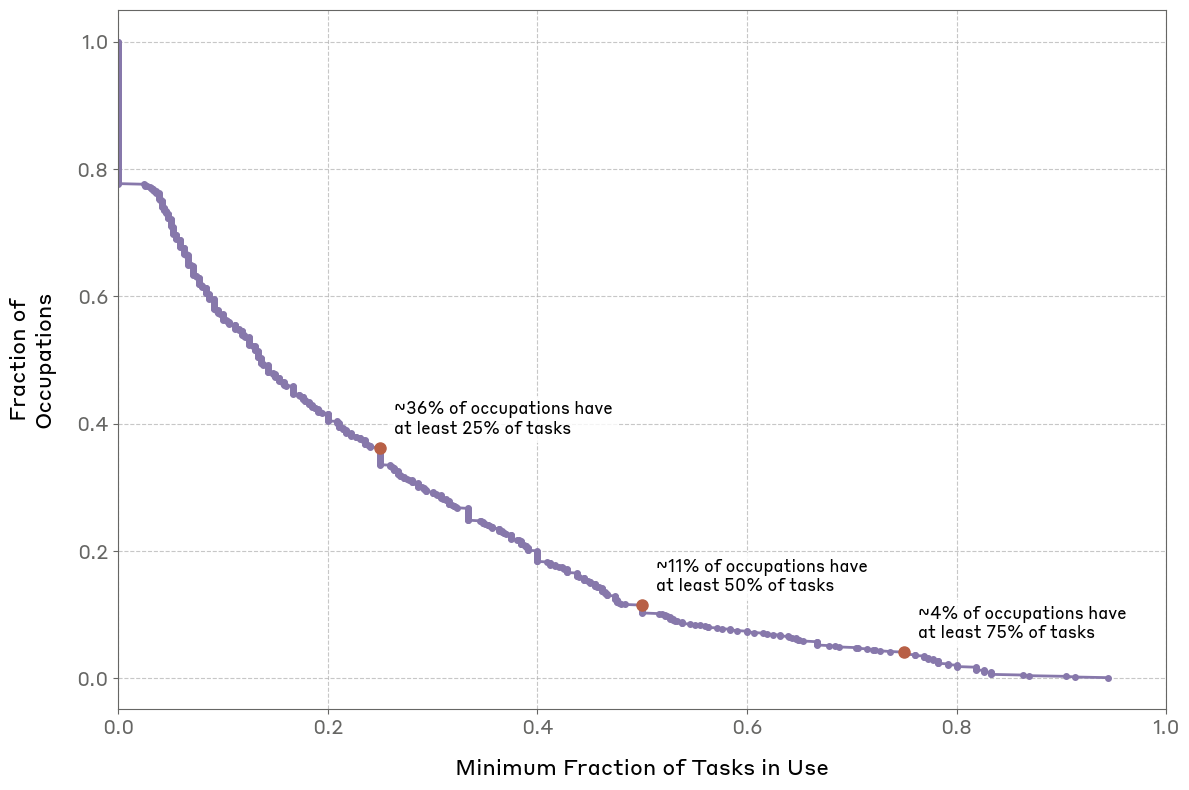}
    \caption{\textbf{Depth of AI usage across occupations.} Cumulative distribution showing what fraction of occupations (y-axis) have at least a given fraction of their tasks with AI usage (x-axis). Task usage is defined as occurrence across five or more unique user accounts and fifteen or more conversations. Key points on the curve highlight that while many occupations see some AI usage ($\sim\!36\%$ have at least 25\% of tasks), few occupations exhibit widespread usage of AI across their tasks (only $\sim\!4\%$  have 75\% or more tasks), suggesting AI integration remains selective rather than comprehensive within most occupations.}    \label{fig:fraction_tasks_adopted}
\end{figure}

\paragraph{How many tasks per occupation show AI usage?} To assess the depth of AI integration across occupations, we examined what fraction of each occupation's tasks appeared in our Clio run.\footnote{Clio operates with a set of privacy measures that protect the activity of individual users from appearing in Clio outputs. In this work, we only examine tasks that have accumulated at least 15 separate conversations spread across a minimum of five different user accounts---ensuring that a task must appear multiple times across different users to appear in our dataset. For more information on our Clio's privacy controls, please see \citet{clio}.} As shown in \Cref{fig:fraction_tasks_adopted}, we find that AI task use follows a heavily skewed distribution. Only $\sim\!4\%$ of occupations show usage for at least 75\% of their associated tasks---for example, in Foreign Language and Literature Teachers (75\% of tasks), we observe AI usage for tasks related to collaborating with colleagues on teaching issues and planning course content, though not for activities like writing grant proposals or maintaining student records. Just $\sim\!11\%$ of occupations show usage for half or more of their tasks, as illustrated by Marketing Managers (50\% of tasks) where we see AI usage in tasks involving market research analysis and strategy development, but not in activities like product specification consultation or trade show coordination. The picture broadens at lower thresholds, where $\sim\!36\%$ of occupations see at least a quarter of their tasks with some usage of AI---exemplified by Physical Therapists (25\% of tasks), where we observe AI interactions in tasks related to research and patient education, but not in hands-on treatment or home care instruction. This distribution suggests that while AI could be touching many occupations today, deep integration across most tasks within any given occupation remains rare for now. Rather than completely automating entire job roles, present-day AI appears to be primarily used for specific tasks within occupations.

\subsection{Demonstration of occupational skills}
\label{sec:skills_usage}
\begin{figure*}
    \centering
    \includegraphics[width=0.99\textwidth]{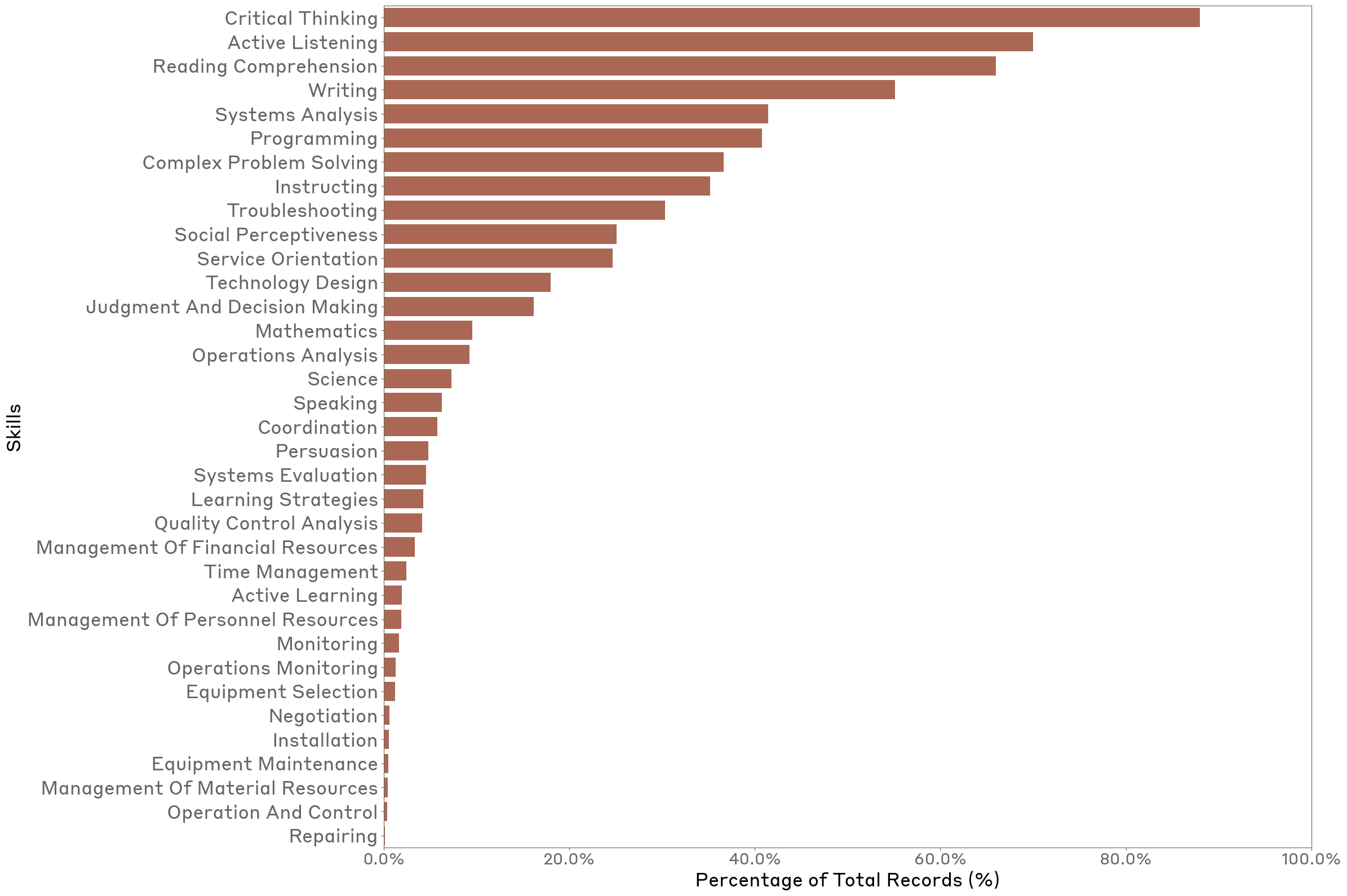}
    \caption{\textbf{Distribution of occupational skills exhibited by Claude in conversations.} Skills like critical thinking, writing, and programming have high presence in AI conversations, while manual skills like equipment maintenance and installation are uncommon.}
    \label{fig:skills_usage}
\end{figure*}

The O*NET database contains 35 occupational skills that identify the essential abilities workers need in order to perform tasks across different jobs. These include skills such as Critical Thinking, Complex Problem Solving, Persuasion, and Equipment Maintenance. We use Clio to identify all of the occupational skills exhibited by the moel in relevant to a given Claude.ai conversation, shown in \Cref{fig:skills_usage}.

Intuitively, skills requiring physical interaction, including Installation, Equipment Maintenance, and Repairing, showed the lowest prevalence in Claude.ai traffic. By contrast, cognitive skills such as Critical Thinking, Reading Comprehension, Programming, and Writing had the highest prevalence. However, our analysis captures only whether a skill was exhibited in Claude's responses, not whether that skill was central to the user's purpose or was performed at an expert level. For instance, while Active Listening appears as the second most common skill, this likely reflects Claude's default conversational behaviors---such as rephrasing user inputs and asking clarifying questions---rather than users specifically seeking out listening-focused interactions.

\begin{figure*}
    \centering
    \includegraphics[width=0.99\textwidth]{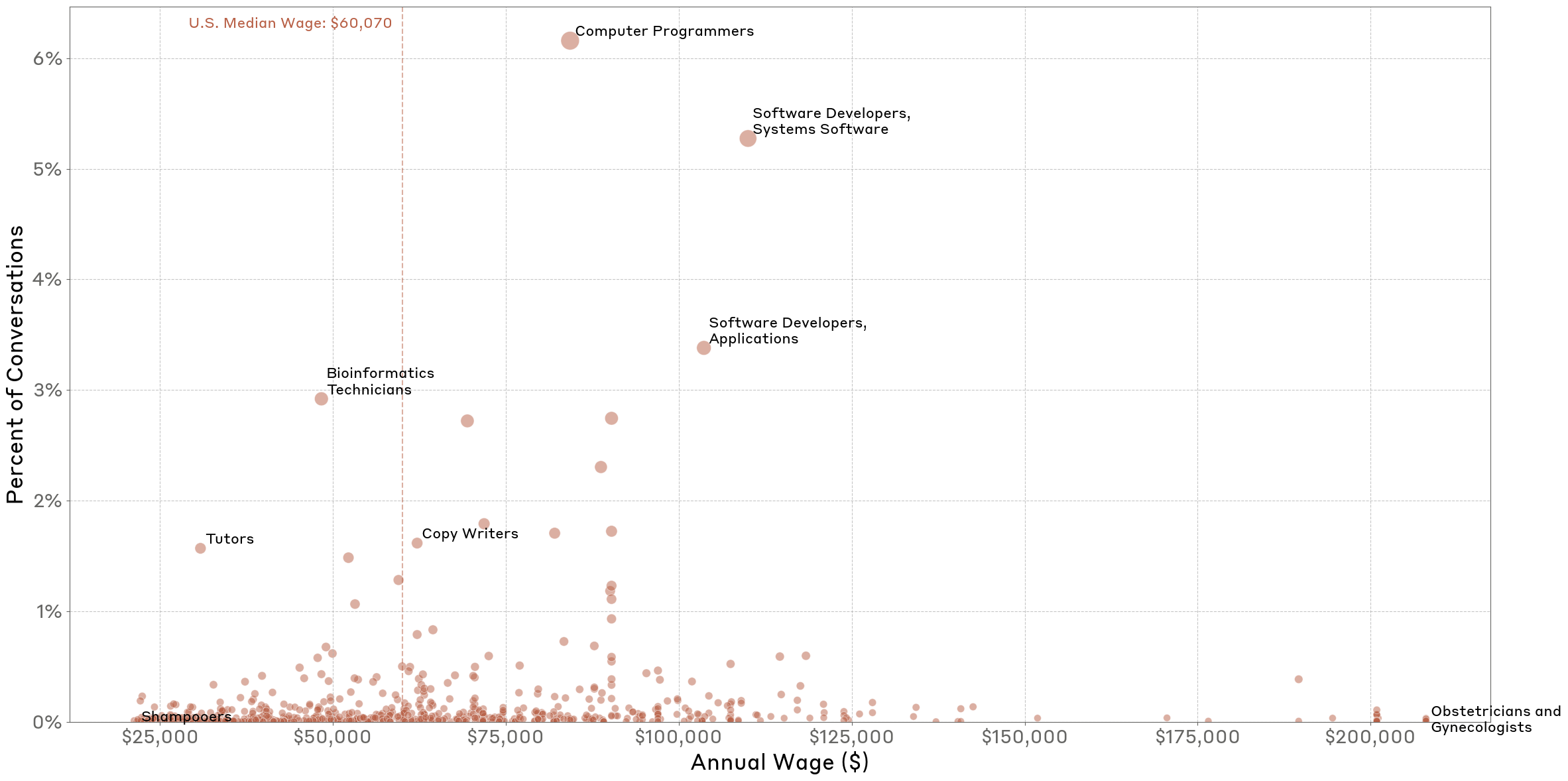}
    \caption{\textbf{Occupational usage of Claude.ai by annual wage.} The analysis reveals notable outliers among mid-to-high wage professions, particularly Computer Programmers and Software Developers. Both the lowest and highest wage percentiles show substantially lower usage rates. Overall, usage peaks in occupations within the upper wage quartile, as measured against U.S. median wages \citep{uscensus2022income}.}
    \label{fig:wage_by_usage}
\end{figure*}

\subsection{AI usage by wage and barrier to entry}
\label{sec:usage_by_wage}
We also report trends in usage across two additional occupational dimensions present in O*NET: the median wage of an occupation and its barrier to entry (i.e., the level of preparation needed for an occupation).   

\paragraph{Wage}
\Cref{fig:wage_by_usage} shows how usage of AI varies by the median wage of that occupation. We find that usage peaks in the upper quartile of wages with computational occupations such as Computer Programmers and Web Developers. Occupations at both extremes of the wage scale show lower usage. For example, waiters and anesthesiologists (low- and high-wage occupations, respectively) are among the least represented in the data, which is in line with our findings that skills requiring physical interaction are least common in our data. 

\paragraph{Barrier to entry}
Occupations in the O*NET database are placed in Job Zones ranging from one to five, categories defined by the amount of preparation needed for a human to perform the duties of a given occupation. Occupations requiring higher levels of education, experience, and training are placed in higher Job Zones. As Job Zone increases from one to four, so does the representation of that zone in our data, peaking at Zone 4: Considerable Preparation Needed, which comprises occupations that typically require a four-year bachelor's degree. However, representation falls for Job Zone 5: Extensive Preparation Needed, where most occupations require advanced degrees. These results align with our wage analysis which showed high usage in the upper quartile of jobs (e.g., software developers) but low usage at the highest end of the spectrum (e.g., anesthesiologists and gynecologists), which tend to require advanced training. These results make clear that human barriers to entry may be significantly different than barriers to language models. See \Cref{appendix:barrier_to_entry,tab:barrier_to_entry} for the complete results. 

\subsection{Automating vs. augmenting users}
\label{sec:autaug}
While the previous analyses reveal which tasks are seeing the most AI usage, they do not tell us \textit{how} AI is being used for these tasks. A key distinction in the economics literature is between automation---where technology substitutes for human labor---and augmentation---where technology complements and enhances human capabilities \citep{autor2015there}. To understand which pattern is more prevalent, we used Clio to classify conversations into one of five different collaboration patterns\footnote{We conduct human validation of these classifications. Details are included in \Cref{appendix:validation}.}
grouped into \textit{automative} vs. \textit{augmentative} behaviors, listed in \Cref{tab:collaboration-patterns}.

\begin{table*}[th]
\begin{tabular}{p{0.48\textwidth}p{0.48\textwidth}}
\toprule
\multicolumn{1}{l}{\textbf{Automative Behaviors}} & \multicolumn{1}{l}{\textbf{Augmentative Behaviors}} \\
\textit{AI directly executes tasks with minimal human involvement} & \textit{AI enhances human capabilities through collaboration} \\
\midrule
\textbf{Directive:} Complete task delegation with minimal interaction & 
\textbf{Task Iteration:} Collaborative refinement process \\[0.2em]
\small\textit{Illustrative Example: ``Format this technical documentation in Markdown''} & 
\small\textit{Illustrative Example: ``Let's draft a marketing strategy for our new product. ... Good start, but can we add some concrete metrics?''} \\[1em]
\textbf{Feedback Loop:} Task completion guided by environmental feedback & 
\textbf{Learning:} Knowledge acquisition and understanding \\[0.2em]
\small\textit{Illustrative Example: ``Here's my Python script for data analysis -- it's giving an IndexError. Can you help fix it? ... Now I'm getting a different error...''} &
\small\textit{Illustrative Example: ``Can you explain how neural networks work?''} \\[1em]
&
\textbf{Validation:} Work verification and improvement \\[0.2em]
&
\small\textit{Illustrative Example: ``I've written this SQL query to find duplicate customer records. Can you check if my logic is correct and suggest any improvements?''} \\
\bottomrule
\end{tabular}
\caption{\textbf{Taxonomy of Human-AI Collaboration Patterns.} We classify conversations into five distinct patterns across two broad categories based on how people integrate AI into their workflow.}
\label{tab:collaboration-patterns}
\end{table*}

Both augmentative and automative collaboration behaviors are present in users' with Claude.ai interactions, with slightly more conversations being labeled as augmentative (57\%) than automative (43\%). That said, we note an important caveat that users might edit and adjust the response they receive from Claude outside the chat window, suggesting that the true proportion of augmentative conversations may be even higher. In addition, even automation of simple tasks can serve to enhance human capabilities when embedded within broader human-directed workflows.

To better understand the distribution of tasks across each of these collaboration patterns, we consider how automative versus augmentative behaviors differ across different occupational tasks:

\paragraph{Automative behaviors}
A majority of the Directive conversations consisted of writing and other content generation tasks. There was also a high percentage of Directive conversations in the business-relevant tasks such as “Draft and optimize professional business email communications” and schoolwork-related tasks such as “Solve diverse geometry problems with calculations and proofs.” The vast majority of the Feedback Loop conversations were related to coding and debugging, where the user repeatedly relayed the error they received back to the model.

\paragraph{Augmentative behaviors}
Task Iteration conversations often involved front-end development (tasks such as "Assist with web development tasks and UI improvements" and "Create and modify landing pages and key website components") as well as professional communication tasks (for example, "Optimize resumes, cover letters, and job applications" and "Assist with professional and academic writing and communication"). The highest percentage of Learning conversations occurs in general education tasks such as “Explain and analyze martial law implementation and impacts,” “Offer gastroenterology and digestive health advice,” and “Assist with microcontroller programming and embedded systems projects”. Validation was the smallest category of conversations, and was nearly all concentrated to tasks discussing language translations.

\subsection{Usage patterns by model type}
\label{sec:usage_by_model}

As AI capabilities evolve, understanding how different models are used can help anticipate shifts in usage across occupations. We explore this by comparing usage patterns between two Claude models: Claude 3 Opus, released in March 2024, and Claude 3.5 Sonnet (new), released in October 2024.

\begin{figure*}
    \centering
    \includegraphics[width=0.99\textwidth]{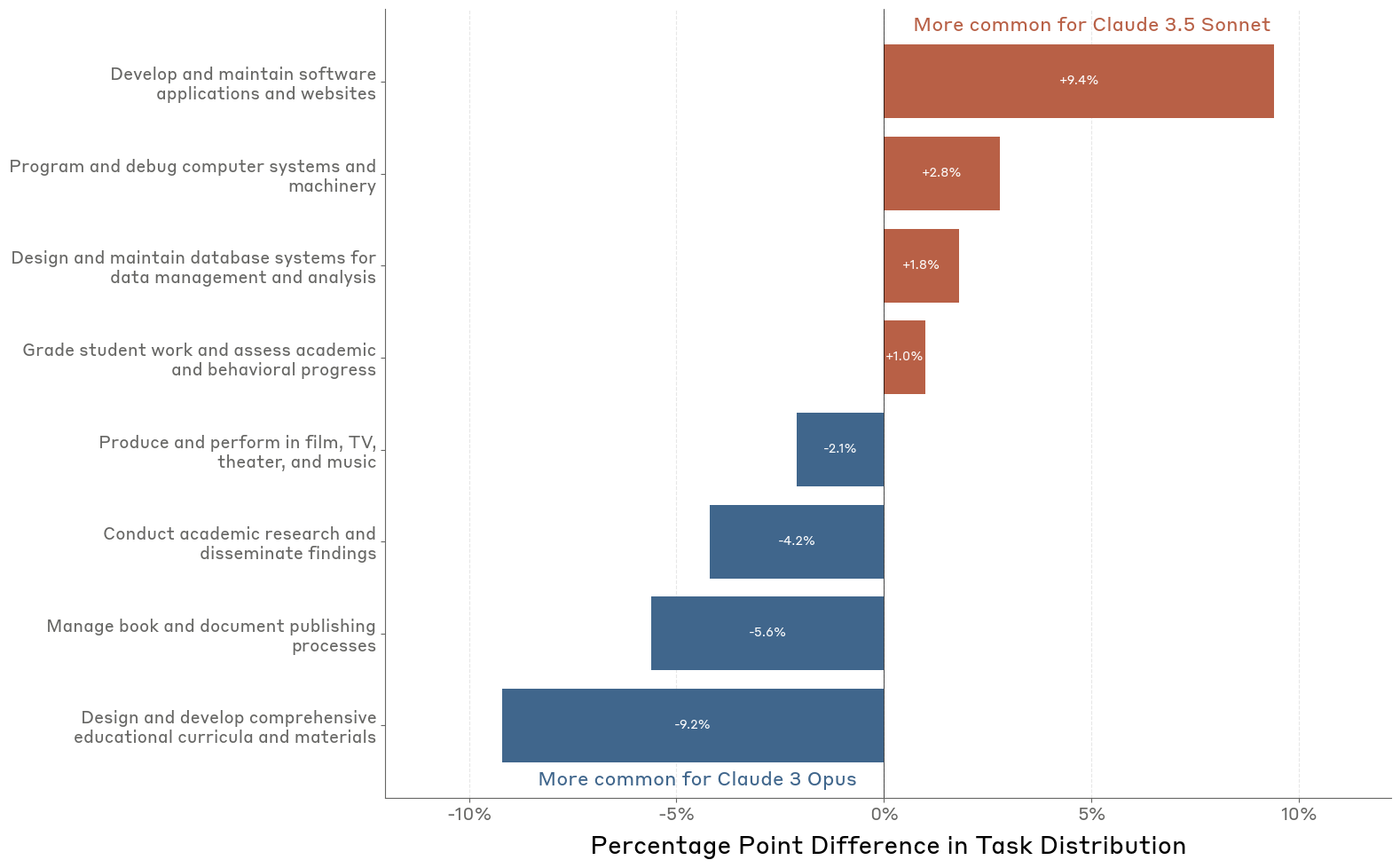}
    \caption{\textbf{Comparative analysis of task usage patterns between Claude Sonnet 3.5 (New) and Claude Opus models,} showing differential preferences in usage. Sonnet 3.5 (New) demonstrates more usage for coding and technical tasks, while Opus is more used for creative writing and educational content development.}
    \label{fig:adoption_by_model}
\end{figure*}

Our analysis reveals clear specialization in how these models are used (\Cref{fig:adoption_by_model}). Relative to Sonnet, Opus sees higher usage for creative and educational work (e.g., "Produce and perform in film, TV, theater, and music," "Manage book and document publishing process,"  "Design and develop comprehensive educational curricula and materials," "Conduct academic research and disseminate findings"). These patterns align with widespread user observations about Opus' relatively unique character and writing style.\footnote{\href{https://www.anthropic.com/research/claude-character}{https://www.anthropic.com/research/claude-character}} By contrast, Claude 3.5 Sonnet (new) is preferred for coding and software development tasks (e.g., "Develop and maintain software applications and websites" and "Program and debug computer systems and machinery"), consistent with external evaluations highlighting its relatively strong coding abilities.

By tracking these usage patterns across model versions at a task-level, we can better understand which capability improvements drive meaningful changes in AI usage across different sectors of the economy.

\section{Discussion}

We present the first large-scale empirical analysis of how advanced AI systems are actually being used across economic tasks. While our work offers broad insights on AI's use in the economy, we note key limitations and areas for future research. 

\subsection{Limitations}
\label{sec:limitations}

\paragraph{Data sample} We use snapshots of Claude.ai Free and Pro conversations over 7-day periods.\footnote{See \Cref{appendix:metadata} for exact date ranges.} It is possible that this sample is not representative of usage on Claude.ai across longer time windows, and quite likely that our sample differs in important ways from API data or data from other AI model providers due to differing model capabilities, product features, and user bases. Additionally, Claude.ai only outputs text as opposed to other modalities. This removes key potential users who may rely on image or video outputs (e.g. fashion designers). For these reasons our findings should be interpreted as an imperfect snapshot of AI usage across the labor market, while noting that a broader understanding of model interaction patterns will emerge as more researchers and organizations are able to share usage data from diverse deployment contexts.

\paragraph{Reliability of model-driven classification} Our use of Claude to classify user conversations may also introduce potential inconsistencies if the model's understanding of tasks differs from the intended reading in the O*NET database. While we conduct human validation of our pipeline (\Cref{appendix:validation}), rely upon past validations of Clio \citep{clio}, and corroborate our results with cluster-level analyses (\Cref{appendix:cluster_experiments}), it is important to note that these classifications likely contain some inherent noise.

\paragraph{Varying complexity of users' queries} 
While we make efforts to exclude conversations without relevance to any occupational tasks (\Cref{appendix:experimental_details}), our method does not account for the complexity of user queries---for example, providing instructions for a basic omelette does not indicate culinary expertise. Consequently, we may overestimate usage rates for certain tasks by classifying conversations from novice users.

\paragraph{Limitations of the O*NET database} While the O*NET database offers valuable insights into current economic sectors, its static nature presents key limitations for analyzing AI's impact on the labor market. The database cannot capture emerging tasks and occupations that AI systems such as Claude may create or transform. Additionally, while O*NET captures a very large number of tasks, it cannot contain all tasks in the economy. Furthermore, these tasks are often written in general terms, leading to inherent ambiguity when classifying conversations---many tasks are similar across multiple different occupations. Finally, as a U.S.-centric classification system, O*NET may overlook significant occupational categories and tasks from other regions, potentially skewing our distributional analysis of global Claude.ai usage. This limits our analysis as AI usage patterns can vary across international contexts \citet{Gmyrek2023GenerativeAA}. 

\paragraph{Lack of full context into user workflows} Although our work analyzes conversation data on Claude.ai, our methods are not able to capture \textit{how} users are using the outputs of Claude.ai conversations. For example, we cannot observe whether users are copying code snippets into development environments, incorporating writing suggestions into documents, fact-checking responses against other sources, or using outputs as inspiration rather than verbatim content. Thus, it remains out of reach to make definitive judgments as to how much Claude's outputs are actually incorporated by users in their tasks. We aim to provide the preliminary framework and findings for this further study to take place.

\subsection{Implications and future work}

While acknowledging these limitations, our analysis reveals several key implications for how we study and respond to AI's economic effects.

\paragraph{Comparison to predictive studies} Our empirical findings both validate and challenge previous predictions about AI's impact on work. \citet{Webb2019TheIO} predicted highest AI exposure in occupations around the 90th wage percentile, while we find peak usage in mid-to-high wage occupations, with notably lower usage at both extremes of the wage distribution. This pattern suggests that factors beyond technical feasibility---such as implementation costs, regulatory barriers, and organizational readiness---may be tempering adoption in the highest-wage sectors. \citet{gptsaregpts} predicted that 80\% of U.S. workers could have at least 10\% of their work tasks affected by language models; by contrast, our empirical data shows current adoption at $\sim\!57\%$ of occupations using AI for at least 10\% of their tasks---lower than predicted but potentially trending toward their forecast as capabilities improve and adoption barriers decrease. However, their prediction of higher usage in industries like healthcare has not yet materialized in our data, and we observe noticeably higher usage in scientific applications than they expected, highlighting the impact of both sector-specific barriers to diffusion as well as unexpected developments in model capabilities. These discrepancies between predictions and actual usage underscore the importance of empirical measurement in understanding AI's evolving economic impact and suggest that technical feasibility alone may not be sufficient to predict where and how AI will be adopted across the economy.

\paragraph{Dynamic tracking of AI usage} Our research provides a framework for systematically tracking AI's integration into the workforce over time. Unlike surveys that capture self-reported behavior, our approach reveals actual AI usage patterns as they naturally occur in the workplace, providing a more accurate and granular picture of true integration. This ability serves multiple crucial functions: it enables early detection of emerging usage patterns, helps identify sectors approaching technological inflection points, and reveals where adoption barriers may be creating uneven diffusion across industries. By monitoring both the breadth (across occupations) and depth (within specific roles) of AI usage, policymakers can develop targeted interventions---whether supporting sectors showing promising (or lagging) productivity gains or addressing potential displacement effects in areas of rapid automation. A dynamic measurement system provides critical lead time for policymakers and organizations to prepare for technological transitions, rather than responding reactively after disruptions have already occurred.

\paragraph{Task-level measurement} Our findings highlight the importance of analyzing AI use at the task level rather than at the  job level. Currently, we observe usage concentrated in specific tasks (e.g., software engineering, content creation) rather than wholesale automation of occupations. If this pattern persists---with AI affecting only a subset of tasks within jobs---it suggests occupations will evolve rather than disappear. However, if the breadth of task usage grows without signs of saturation, that may suggest the possibility of a more comprehensive workplace transition.

\paragraph{Augmentation vs automation} Within affected tasks, the way AI systems are used can differ significantly. Our analysis reveals an important distinction: while some users employ AI systems to completely automate tasks, others use them as collaborative tools that enhance their capabilities. This difference matters for both workers and productivity. When AI serves as an augmentative partner rather than a replacement, studies have shown improved productivity while maintaining individuals' meaningful engagement in their work \citep{noy2023experimental, peng2023impact, cui2024effects}. These patterns can inform policy priorities---supporting the development of collaborative AI interfaces where they show clear benefits, while ensuring appropriate preparation for areas where automation becomes more prevalent.

\paragraph{From usage patterns to broader impacts}
Understanding how current AI usage patterns may translate into broader economic changes remains a key challenge. While our data reveals where AI is being used today, inferring long-term consequences from these early usage trends poses significant empirical challenges \citep{Acemoglu2022ArtificialIA}. For instance, high usage in certain occupations could signal future productivity gains or displacement effects, while the uneven distribution of AI use across wage levels may offer early indicators of how AI could reshape economic opportunities and inequalities. While our present results cannot definitively map these relationships, longitudinal analysis tracking both usage patterns and outcomes could help reveal the mechanisms by which AI usage drives changes in the workplace.

\medskip
Overall, our findings demonstrate that AI has already begun to see use across a significant fraction of economic tasks. We offer this initial framework for tracking AI's evolving impact on work as a starting point, and hope to partner with policymakers, economists, and other stakeholders to develop policy proposals that will spread the benefits of AI across the economy.

\section{Conclusion}

Understanding how AI affects the economy requires grounding our analysis in real-world data. Our analysis of millions of conversations on Claude.ai reveals clear patterns: AI usage peaks in software development and technical writing, with $\sim\!4\%$  of occupations showing usage across three-quarters of their tasks and $\sim\!36\%$  showing usage in at least a quarter of their tasks. Usage splits nearly evenly between automation (43\%) and augmentation (57\%) of human capabilities. While these patterns are informative, they capture just the beginning of AI's integration into work. As AI systems expand beyond text to handle video, speech, and physical actions through robotics, and as AI agents become more capable of carrying out extended tasks autonomously, the nature of human-AI collaboration is poised to transform dramatically. New tasks and even entirely new occupations may emerge around these capabilities. Empirical frameworks that track these changes dynamically will be crucial for anticipating and preparing for the evolving landscape of work. The challenge ahead lies not just in measuring these changes, but in using our understanding of them to help shape a better future.

\section{Acknowledgements}

We thank Avital Balwit, Landon Goldberg, Logan Graham, Zac Hatfield-Dodds, Andrew Ho, Kamya Jagadish, Rebecca Lee, Liane Lovitt, Jennifer Martinez, Andi Peng, Ankur Rathi, Orowa Sikder, Colt Steele, Janel Thamkul, and Meg Tong for their helpful ideas, discussion, and feedback. Additionally, we appreciate the productive comments and discussion from Jonathon Hazell, Anders Humlum, Molly Kinder, Anton Korinek, Benjamin Krause, Michael Kremer, John List, Ethan Mollick, Lilach Mollick, Arjun Ramani, Will Rinehart, Robert Seamans, Michael Webb, and Chenzi Xu on early findings and drafts of the paper.

\bibliography{custom}

\appendix
\label{sec:appendix}

\newpage

\section{Author Contributions}
\label{appendix:contributions}

\textbf{Kunal Handa} led experimentation and implementation, including core contributions to methodology and analysis. 
\textbf{Alex Tamkin} proposed the idea and led the design of the methods and analysis.
\textbf{Miles McCain} built the underlying technical infrastructure, partnered with Kunal Handa to develop the task hierarchy and assignment system, led the cluster reconstruction analysis, and supported all experiments.
\textbf{Saffron Huang} and \textbf{Esin Durmus} contributed key feedback across the project and assisted with the related work section and validation.
\textbf{Dario Amodei}, \textbf{Jared Kaplan}, and \textbf{Jack Clark} provided valuable guidance, support, and discussion throughout the process.
\textbf{Deep Ganguli} provided detailed guidance, organizational support, and feedback throughout all stages of the project—experiments, technical infrastructure, analysis, validation, and writing.
\textbf{All other authors} contributed to the framing, experiments, analysis, figures, or other efforts that made our work possible.

\section{Methodological Details}
\label{appendix:experimental_details}
To conduct our analysis, we use Clio \citep{clio}, a privacy-preserving framework for analyzing language model conversations. By leveraging AI assistants themselves to surface and analyze aggregated usage patterns across millions of conversations, Clio enables us to systematically study human-AI conversations in the real-world. Clio uses \textit{facets} to extract specific attributes or characteristics from conversations. Facets can capture both direct questions like "What task is the assistant performing?" as well as metadata such as the AI model used to fulfill a request. In this work, we use Clio to move beyond theoretical predictions to understand which occupational tasks AI systems are actually being used for as well as to characterize the nature of these interactions. 

\subsection{Usage Across Tasks and Occupations} \label{appendix:adoption_across_tasks}

\begin{figure}[h]
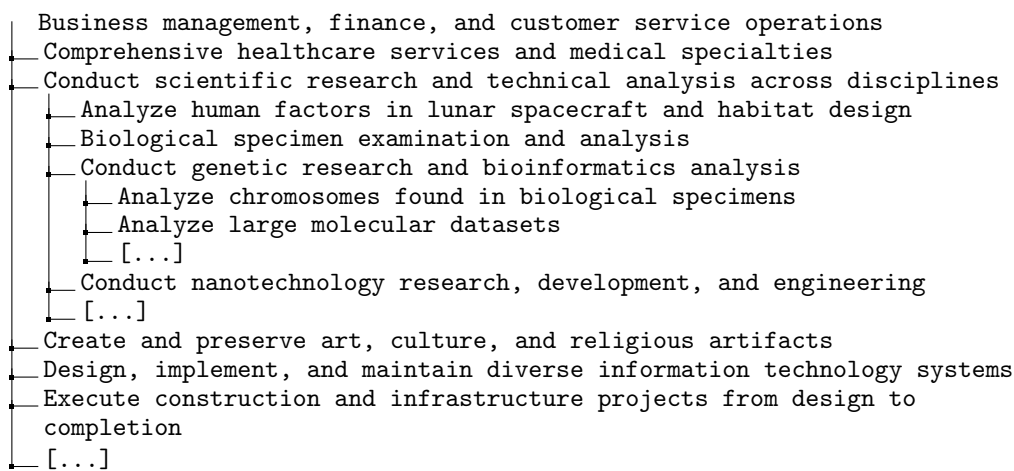

\dirtree{%
.1 Business management, finance, and customer service operations.
.1 Comprehensive healthcare services and medical specialties.
.1 Conduct scientific research and technical analysis across disciplines.
.2 Analyze human factors in lunar spacecraft and habitat design.
.2 Biological specimen examination and analysis.
.2 Conduct genetic research and bioinformatics analysis.
.3 Analyze chromosomes found in biological specimens.
.3 Analyze large molecular datasets.
.3 [...].
.2 Conduct nanotechnology research, development, and engineering.
.2 [...].
.1 Create and preserve art, culture, and religious artifacts.
.1 Design, implement, and maintain diverse information technology systems.
.1 Execute construction and infrastructure projects from design to completion.
.1 [...].
}
\caption{Example subsection of the generated O*NET task hierarchy. Our hierarchy contains three levels: 12 top-level tasks, 474 middle-level tasks, and 19530 base-level (O*NET) tasks.  }
\label{fig:task-hierarchy}
\end{figure}

With Clio, we map Claude.ai conversations to specific occupational tasks and their associated characteristics. However, a key challenge is the size of the O*NET task database we use ($\sim\!20$K tasks), which makes direct classification via zero or few-shot prompting impossible because the full list of tasks does not fit in the model's context window. We instead construct this as a classification over a hierarchy of task labels (\Cref{fig:task-hierarchy}), inspired by \citet{hierarchicalMorin, hierarchicalMnih}.

Our approach consists of three main components: 1) creating a hierarchical taxonomy of the O*NET occupational tasks, 2) mapping from summarized, privacy-preserved, conversations to O*NET tasks with Clio, 3) connecting O*NET tasks to occupations.

We report the specific number of conversations used for each analysis and other relevant metadata in \Cref{appendix:metadata}.

\paragraph{Creating a task hierarchy}
\label{appendix:task-hierarchy}
The O*NET database contains $\sim\!20$K task descriptions across all occupations. We constructed a multi-level taxonomy of tasks using Clio's hierarchy generation step (\Cref{fig:task-hierarchy}). This process recursively organizes base-level tasks into broader categories. The following details are reproduced from Appendix Section G.7 of \cite{clio}.\footnote{The key difference here is that rather than generating clusters from conversations, we are generating clusters from O*NET tasks.} 

We: 
\begin{itemize}
    \item \textbf{Embed task names.} Embeds tasks names using the \textbf{all-mpnet-base-v2} \citep{all_mpnet_base_v2} sentence transformer to obtain 768-dimensional vector representations of each task
    \item \textbf{Generate neighborhoods.} Group these embeddings into $k$ neighborhoods using $k$-means clustering, where $k$ is chosen so that the average number of tasks per neighborhood is 40. We group tasks into neighborhoods because the names and descriptions for all base clusters may not fit within Claude's context window.
    \item \textbf{Propose new tasks for each neighborhood.} For each neighborhood, use Claude to propose candidate higher-level task descriptions by examining both the tasks within the neighborhood and the nearest $m$ tasks outside it. Including the nearest tasks \textit{beyond} the neighborhood ensures that tasks (or groups of tasks) on the boundary between neighborhoods are neither overcounted nor undercounted. We require the final number of tasks at the level $l$ to be $n_l \pm 1.5 n_l$, where $n_l$ is chosen such that the ratio between successive levels follows $n_l/n_{l-1} = (n_{\text{top}}/n_{\text{base}})^{1/(L-1)}$ for $L$ total levels.
    \item \textbf{Deduplicate across neighborhoods.} Deduplicate and refine the proposed tasks across all neighborhoods using Claude to ensure distinctiveness while maintaining coverage of the underlying data distribution.
    \item \textbf{Assign to new best fit higher-level tasks.} Assign each lower-level task to its most appropriate parent task using Claude. We randomly shuffle the order of the higher-level tasks when sampling from Claude to avoid biasing assignments based on the order of the list.
    \item \textbf{Rename higher-level tasks.} Once all tasks at level $l$ have been assigned to a higher-level task, we \textit{regenerate} a new name and description for the parent task based on the lower-level tasks that were assigned to it. This renaming step ensures that task names continue to accurately reflect their contents.
\end{itemize}

\paragraph{Mapping conversations to O*NET tasks}

To map conversations to specific tasks, we use this generated hierarchy to perform a tree-based search through our task hierarchy. For each conversation, we first used Claude to determine if the conversation was occupationally relevant. We screen conversations using Claude 3.5 Haiku (\texttt{claude-3-5-haiku-20241022}). The prompt we use for screening is provided in \Cref{appendix:prompts}. If the conversation is deemed relevant, we traverse the hierarchy from top to bottom, with Claude selecting the most appropriate task at each level based on the conversation content. Through this process, Clio calculates the cumulative number of conversations assigned to each task in the O*NET database. For privacy reasons, tasks with less than 5 unique accounts or 15 conversations are excluded from our analysis. This also serves to reduce statistical noise. Complete prompts are included in \Cref{appendix:prompts}. We also experimented with multi-class classification allowing up to k=3 task assignments per conversation; results were qualitatively similar, so we report the single-class results here for simplicity. 
 
\paragraph{Connecting O*NET tasks to occupations}
Each task in O*NET is associated with one or more occupations. To conduct our occupation-level analysis, we map from individual tasks to their associated occupations. To calculate occupational use, we aggregate the number of conversations associated with tasks for a given occupation. In the case that a single task maps to multiple occupations, each occupation's conversation count is incremented by the number of conversations assigned to that task divided by the number of occupations associated with that task.

\subsection{Conversation-level vs. Account-level Analysis}
\label{appendix:conversations-vs-accounts}
\begin{figure*}
    \centering
    \includegraphics[width=0.99\textwidth]{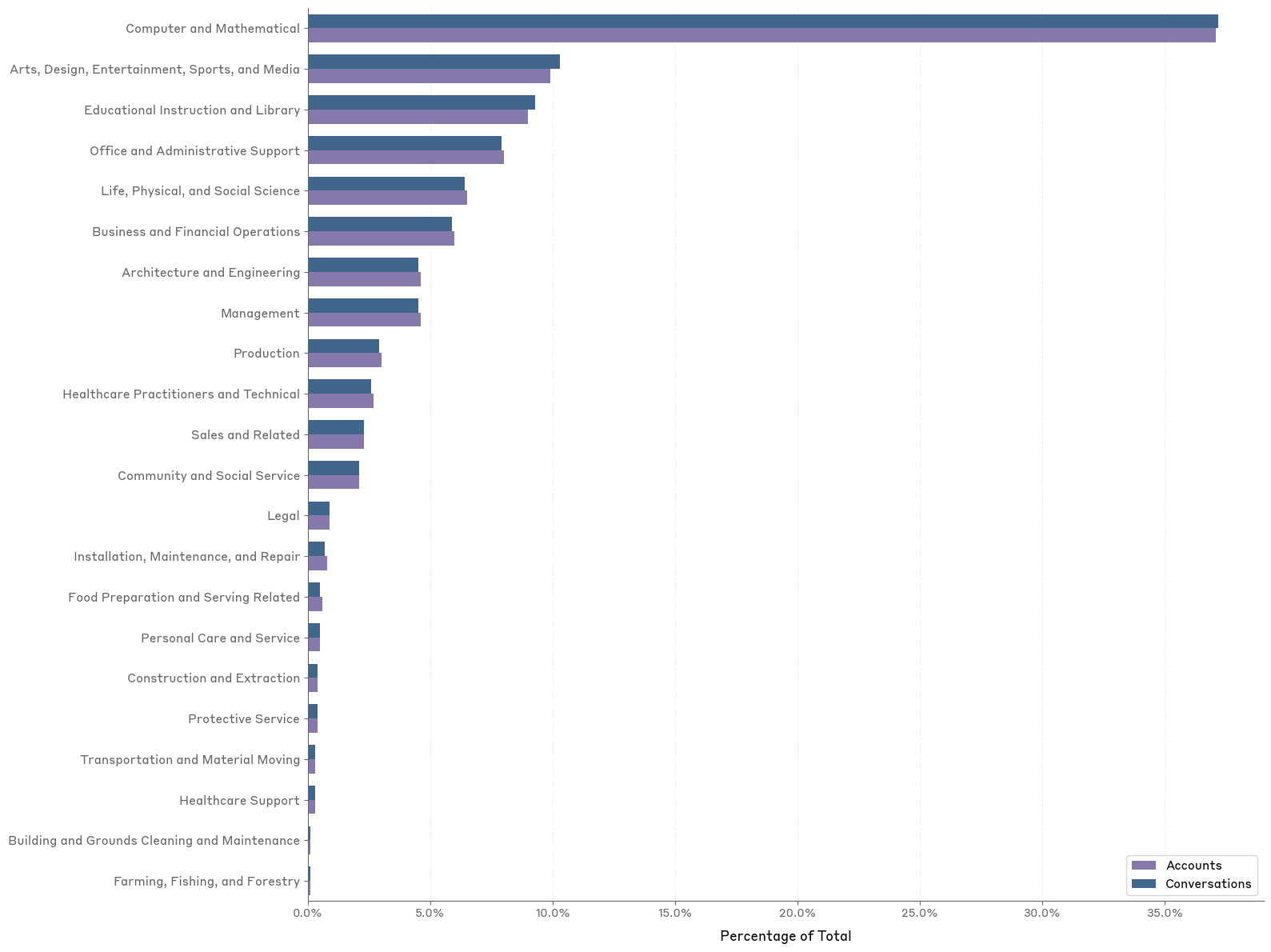}
    \caption{We observe minimal difference in our measurements of AI use across occupational categories when measuring by number of conversations versus number of accounts}
    \label{fig:convos_vs_accounts}
\end{figure*}

To assess the robustness of our methodology for measuring AI usage across occupations, we examined if our focus on total conversation counts could systematically bias results toward occupations that tend to generate many short interactions (e.g. programmers debugging code through multiple separate conversations). To address this, we compared two different approaches to analyze our task-level classifications: measuring the total number of conversations per task (as discussed in the main text) vs. measuring the unique user accounts per task.  

As shown in \Cref{fig:convos_vs_accounts}, the patterns of AI usage across occupations remain remarkably stable regardless of which approach we use. This suggests our findings are not affected by variation in interaction patterns across occupations.

\subsection{Usage by Wage and Barrier to Entry}
\label{appendix:usage_by_wage}
\paragraph{Wage} Occupations in O*NET contain associated data on the median wage of workers in that occupation.\footnote{This information is included on the O*NET website but at the time of writing this paper, we did not find it readily available in a downloadable format. We instead rely on data from \citet{kilbournequirk_onetdataviz_2019}, which scraped the O*NET website to obtain this occupation-specific wage data.} To analyze occupational usage of AI relative to occupations' wage, we mapped each occupation's conversation count to the median salary of that occupation. 

\paragraph{Barrier to entry} Occupations in O*NET are placed into one of five Job Zones which describe the amount of preparation a person would need for a given occupation. O*NET delineates the Job Zones as:

\begin{quote}
Job Zone 1 - occupations that need little or no preparation\\
Job Zone 2 - occupations that need some preparation\\
Job Zone 3 - occupations that need medium preparation\\
Job Zone 4 - occupations that need considerable preparation\\
Job Zone 5 - occupations that need extensive preparation
\end{quote}

Occupations are placed into job zones via assessments of their necessary education, related experience, job training, and specific vocational preparation---"the amount of lapsed time required by a typical worker to learn the techniques, acquire the information, and develop the facility needed for average performance in a specific job-worker situation." 

Example occupations for each job zone are:
\begin{quote}
Job Zone 1 - dishwashers, baristas, and agricultural equipment operators\\
Job Zone 2 - customer service representatives, security guards, and tellers\\
Job Zone 3 - electricians, medical assistants, and barbers\\
Job Zone 4 - real estate brokers, sales managers, and database administrators,\\
Job Zone 5 - pharmacists, lawyers, and biologists
\end{quote}

The O*NET website contains further details on requirements for each job zone and example occupations \citep{onet}.

\subsection{Usage by Occupational Skills}
\label{appendix:skills_usage}
The O*NET database contains 35 unique occupational skills which identify key abilities for different occupations \citep{onet}.

Similar to our analyses on \textit{Automating vs. Augmenting Users} and \textit{Usage by Model Type}, we create a Clio facet which assigns conversations to the most relevant occupational skills exhibited by the model. Each conversation is assigned to all skills exhibited by the assistant within the conversation. Assignments to multiple skills are allowed. There is also an option to return None if none of the skills apply to the conversation. In our analyses, we filtered out these classifications as only 0.1\% of conversations were labeled None. For each conversation, the order of skills are shuffled before querying for an assignment to avoid any biases due to ordering. We include the complete prompt in \Cref{appendix:prompts}.

\subsection{Automating vs. Augmenting Tasks}
\label{appendix:augaut}
To understand the nature of users' interactions with Claude, we construct a Clio facet to identify the collaboration pattern present in each conversation. We include our complete prompt in \Cref{appendix:prompts}. We intersect this facet with the previously-mentioned Clio facet mapping from conversations to O*NET tasks. Thus, for each O*NET task, Clio provides an associated breakdown across collaboration patterns. This allows us to conduct fine-grained analysis on the nature of augmentation vs. automation for individual tasks (and thus also for occupations).   

\subsection{Usage by Model Type}
\label{appendix:usage_by_model}
Using Clio, we analyze the distribution of Claude.ai conversations across different model types and O*NET tasks. By combining model usage data with our O*NET task mappings, we examine how Claude 3.5 Sonnet (new) and Claude 3 Opus are utilized across various task categories. Our sample consisted of approximately 54\% Claude 3.5 Sonnet (new) conversations and 46\% Claude 3 Opus conversations.

\subsection{Validating the Composition of our Dataset}
\label{appendix:dataset-composition}

A natural concern with analyzing conversations with an AI assistant like Claude is that usage might be dominated by personal queries, homework help, or casual interactions rather than genuine occupational tasks. To validate that our analysis captures predominantly occupational tasks, we conducted a systematic analysis of conversation types in our dataset. Using Clio, we classified each conversation into one of four categories for three different dimensions: work (explicitly work, likely work, possibly work, non-work), coursework (explicitly coursework, likely coursework, possibly coursework, non-coursework), and personal use (explicitly personal, likely personal, possibly personal, non-personal). We then examined the composition of tasks within each category using Clio.

After applying our occupational task screening, we found that non-work conversations only comprise 23\% of the dataset, and usage relating to coursework comprises only 5-10\% of conversations, depending on the chosen level of confidence. Importantly, we found that the majority of these "non-work" interactions still mapped meaningfully to occupational tasks. For example, personal nutrition planning relates to dietitian tasks, automated trading strategy development connects to financial analyst tasks, and travel itinerary planning maps to travel agent tasks. These results give us increased confidence  in the utility of our dataset for understanding patterns of AI usage across economic tasks, regardless of whether those tasks happen within a formal work context.

\section{Human Validation}
\label{appendix:validation}

We conduct human validation studies on a subset of conversations to assess the quality of our classification framework. To conduct these studies, we used conversations for which users submitted feedback from Claude.ai Free and Pro. We then used Clio to analyze these conversations. Consistent with our Privacy Policy, which allows us to access feedback data for research purposes, we retained the ability to view the underlying user-feedback conversations in the resulting Clio clusters. For more information, please see Anthropic's relevant \href{https://support.anthropic.com/en/articles/8325621-i-would-like-to-input-sensitive-data-into-free-claude-ai-or-claude-pro-who-can-view-my-conversations?q=thumbs}{FAQs} and \href{https://www.anthropic.com/legal/privacy}{privacy policy}. 

While we acknowledge the distribution of conversations for which users submitted feedback likely differs from regular Claude.ai usage, we consider this distribution to be more reliable than alternative open-source datasets.

\paragraph{Task hierarchy} We hand-validate 150 examples for our task hierarchy classifications (\Cref{sec:task_and_occupation_usage} and \Cref{appendix:task-hierarchy}). We examine the classification path traversed by each conversation through our hierarchical framework. We assess the accuracy of labels assigned at each level of the hierarchy:

\begin{itemize}
    \item At the top level, 95.3\% of conversations are judged as assigned to an acceptable task.
    \item At the middle level, 91.3\% of conversations are judged as assigned to an acceptable task.
    \item At the base (O*NET) level, 86\% of conversations are judged as assigned to an acceptable task.
\end{itemize}

As tasks become more specific, classification becomes more difficult. Thus, accuracy of our method predictably decreases as we traverse the hierarchy. Furthermore, there is likely some irreducible noise in the assignment process, given that the O*NET database, while extensive, does not encompass all possible tasks in the economy. As model capabilities increase, we expect classification performance to improve; we treat this as the baseline performance for future efforts to track AI usage using our framework. Although our methods are imperfect, our high validation scores provide confidence in our overall methodology and results. 

\paragraph{Automating vs. augmenting users} We hand-validate 150 examples for our automation vs. augmentation classifications (\Cref{sec:autaug}). We assess whether the label assigned by our framework (Directive, Feedback Loop, Validation, Task Iteration, or Learning) represents the best categorization for each conversation. We find that 90.7\% of conversations are assigned to their optimal label as assessed by human raters. 

\section{Additional Results}

\subsection{Usage Across the Task Hierarchy}
Using Clio we create a task hierarchy to assign conversations to the best fit O*NET task. Here, we build upon our analysis in \Cref{sec:task_and_occupation_usage}, providing results on the most common tasks appearing in Claude.ai data. Our results include the top tasks at each level of our hierarchy. 

\begin{figure*}
    \centering
    \includegraphics[width=0.99\textwidth]{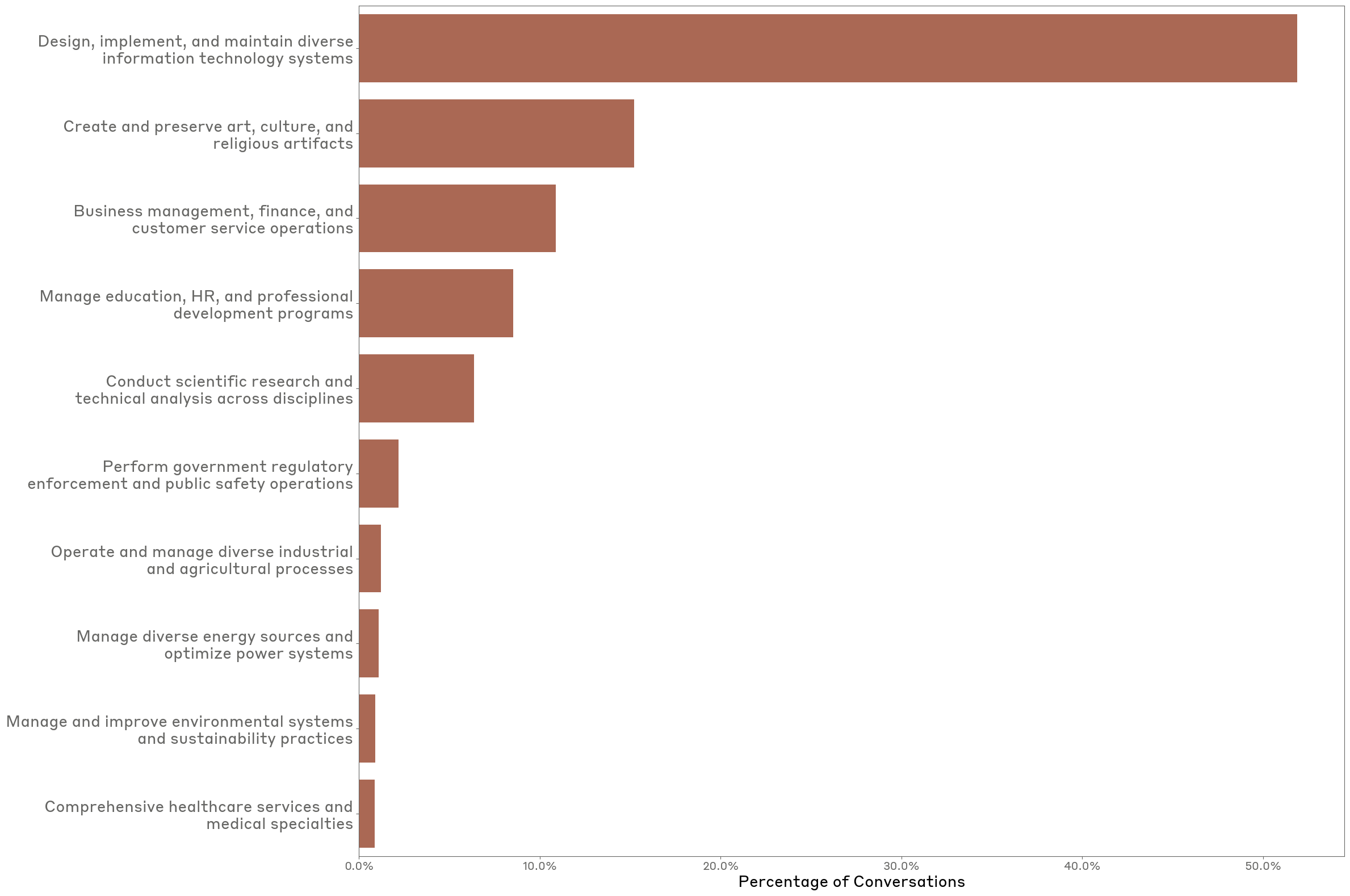}
    \caption{Most prevalent top-level tasks.}
    \label{fig:top_tasks}
\end{figure*}

\textbf{At the top-level} (\Cref{fig:top_tasks}), we see that IT, technology, and associated related tasks dominate the distribution, at nearly 50\% of conversations. The second tier consists of creative and cultural work, with tasks related to creating and preserving art, culture, and religious artifacts making up approximately 20\% of conversations. Business management, finance, and customer service operations form the third major category at around 15\% of conversations. The remaining categories each represent less than 15\% of conversations, with healthcare services and environmental systems showing lower representation at less than 5\% each.

\begin{figure*}
    \centering
    \includegraphics[width=0.99\textwidth]{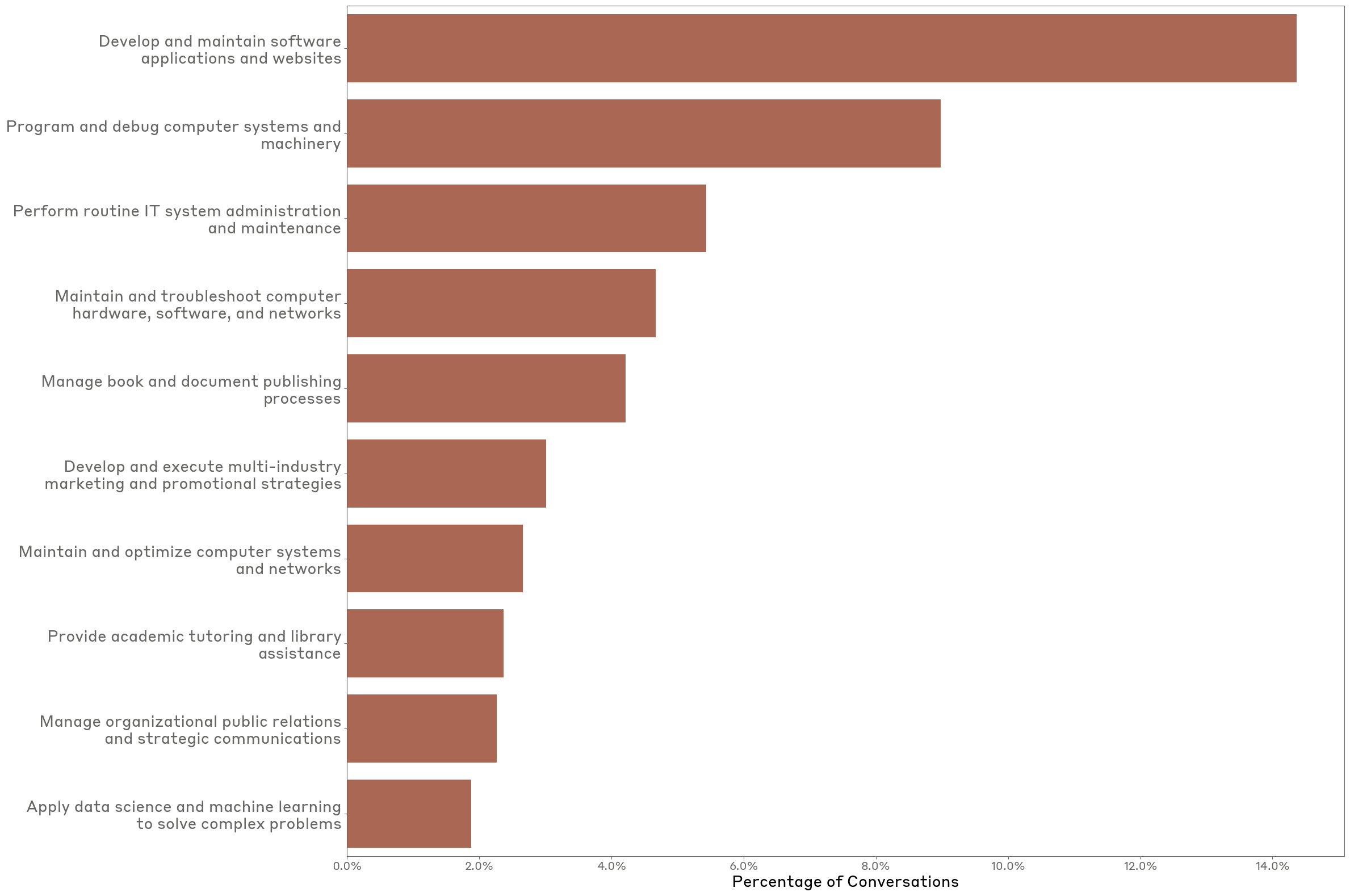}
    \caption{Most prevalent middle-level tasks.}
    \label{fig:mid_tasks}
\end{figure*}

\textbf{At the middle-level} (\Cref{fig:mid_tasks}), the data reveals more granular task patterns. Software development and website maintenance is the most prevalent activity at about 14\% of conversations, followed by computer systems programming and debugging at roughly 11\%. System administration, hardware/software troubleshooting, and document publishing processes each account for 4-6\% of conversations. Marketing/promotional strategies, network optimization, academic tutoring, and public relations management appear but at lower frequencies around 2-3\% each. Data science and machine learning applications represent approximately 2\% of conversations.

\begin{figure*}
    \centering
    \includegraphics[width=0.99\textwidth]{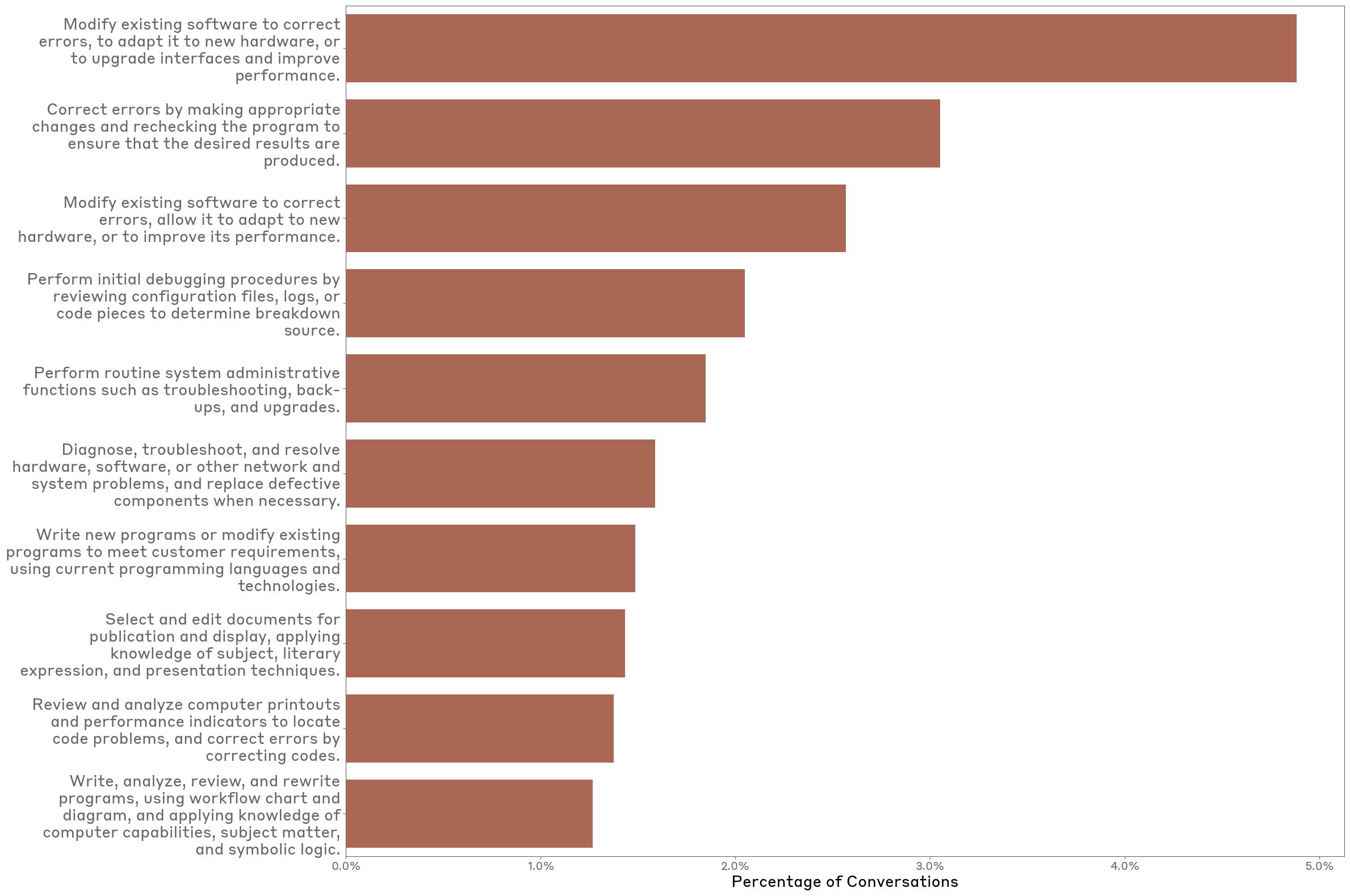}
    \caption{Most prevalent base-level (O*NET) tasks.}
    \label{fig:onet_tasks}
\end{figure*}

\textbf{At the base-level} (O*NET tasks, \Cref{fig:onet_tasks}), we see highly-specific technical operations. Software modification and error correction activities dominate, with tasks focused on adapting software to new hardware or improving performance appearing most frequently. Initial debugging procedures, system administration, and hardware/software troubleshooting follow as the next most common activities. Document editing and program analysis tasks appear less frequently but still constitute a notable portion of conversations.

Our findings suggest that current AI usage is heavily concentrated in technology and content creation roles, with varying degrees of usage across other business functions. For more discussion see \Cref{sec:task_and_occupation_usage}.

\subsection{Barrier to Entry Specifics}
\label{appendix:barrier_to_entry}

\Cref{tab:barrier_to_entry} contains the specific breakdown for our results on AI usage across occupational barriers to entry (\Cref{sec:usage_by_wage}). The data is organized into Job Zones 1-5, representing increasing levels of occupational preparation requirements, from minimal (Zone 1) to extensive preparation (Zone 5). The columns represent:
\begin{itemize}
    \item Conv. \%: The percentage of total usages falling within each zone
    \item Base. \%: The baseline percentage of occupations in each zone
    \item Rep. Ratio: The representation ratio (Conv. \% / Base. \%), indicating relative usage rates
\end{itemize}

The analysis reveals peak AI usage in Job Zone 4 (requiring bachelor's degree-level preparation), with a representation ratio of 1.50, indicating 50\% higher usage than expected given its baseline distribution. Both minimal preparation (Job Zone 1) and extensive preparation roles (Job Zone 5) show lower usage rates, with representation ratios of 0.40 and 0.58 respectively. Job Zones 2 and 3 demonstrate usage rates more closely aligned with their baseline distribution (ratios of 0.78 and 0.94). This pattern suggests AI tools are most readily used in occupations requiring substantial but not elite-level specialization, likely due to these roles involving structured, analytical tasks well-suited to current AI capabilities. The lower usage in both minimal preparation roles and highly specialized positions may occur due to many of those jobs requiring physical manipulation of the environment. For the most specialized tasks, regulations and access to sensitive information (e.g. in healthcare roles) may also play a role in limited AI usage.

\begin{table}
\centering
\begin{tabularx}{\textwidth}{@{}XXXX@{}}
\hline
Job Zone & Conv. \% & Base. \% & Rep. Ratio \\
\hline
1 & 0.23 & 0.57 & 0.40 \\
2 & 9.40 & 12.06 & 0.78 \\
3 & 16.50 & 17.63 & 0.94 \\
4 & 54.45 & 36.33 & 1.50 \\
5 & 19.43 & 33.41 & 0.58 \\
\hline
\end{tabularx}
\caption{\textbf{Analysis of AI usage across occupational barriers to entry,} from Job Zone 1 (minimal preparation required) to Job Zone 5 (extensive preparation required). Shows relative usage rates compared to baseline occupational distribution in the labor market. We see peak usage in Job Zone 4 (requiring considerable preparation like a bachelor's degree), with lower usage in zones requiring minimal or extensive preparation.}
\label{tab:barrier_to_entry}
\end{table}

\section{Analysis Metadata}
\label{appendix:metadata}
Below we provide the metadata for each analysis provided in this paper. We include the number of Claude.ai conversations used and the time span from which these conversations were sampled.

\begin{table}[h]
\begin{tabular}{lll}
\hline
\textbf{Analysis} & \textbf{Conversations} & \textbf{Time Span} \\
\hline
Usage Across Tasks and Occupations (\Cref{sec:task_and_occupation_usage}) & 1M & 2024-12-16 to 2024-12-23 \\
Usage by Wage and Barrier to Entry (\Cref{sec:usage_by_wage}) & 1M & 2024-12-16 to 2024-12-23 \\
Usage by Skills (\Cref{sec:skills_usage}) & 500K & 2025-01-10 to 2025-01-17 \\
Automating vs. Augmenting Users (\Cref{sec:autaug}) & 1M & 2024-12-16 to 2024-12-23 \\
Usage by Model Type (\Cref{sec:usage_by_model}) & 1M & 2024-12-15 to 2025-01-04 \\
\hline
\end{tabular}
\end{table}

\section{Prompts}
\label{appendix:prompts}
Below, we provide our complete prompts for specific analyses and components as outlined above.

\subsection{Screening for Occupationally-Relevant Conversations}
Below is the prompt for the screening process used to determine if a conversation is occupationally relevant as described in \Cref{appendix:experimental_details}.

\begin{spverbatim}
Human: The following is a conversation between Claude, an AI assistant, and a user:

{conversation}

Assistant: I understand.

Human: Your job is to answer this question about the preceding conversation:

<question>
Does the conversation possibly involve an occupational task?
</question>

What is the answer? You MUST answer either only "Yes" or "No". Provide the answer in <answer> tags with no other commentary.

Assistant: Sure, the answer to the question is: <answer> 
\end{spverbatim}

\subsection{Mapping Conversations to O*NET Tasks}
\begin{spverbatim}
Human: The following is a conversation between Claude, an AI assistant, and a user:

{conversation}

Assistant: I understand.

Human: Consider the following list of classification options: 

<options>
{options_str}
</options>
    
Your job is to identify which option best describes the previous conversation. In this case, the provided options are occupational tasks. Your job is to identify which task is performed by the assistant in the previous human-AI assistant conversation.

What is the answer? You MUST provide an option exactly as written above. If multiple options apply, choose the single-most pertinent one. First, start off by considering various aspects of the conversation in <scratchpad> tags in at most four sentences, and then provide the final answer in <answer> tags with no other commentary.

Assistant:
\end{spverbatim}

\subsection{Automating vs. Augmenting Users}
\begin{spverbatim}
Human: Consider the following conversation:

<conversation>
{conversation}
</conversation>

Your task is to analyze human-AI interactions in conversation transcripts to identify the primary collaboration pattern. Focus on how the human structures their requests and engages with the AI assistant. Analyze the interaction according to these collaboration patterns:

- Directive - Human delegates complete task execution to AI with minimal interaction
- Feedback Loop - Human and AI engage in iterative dialogue to complete task with human mainly providing feedback from the environment
- Task Iteration - Human and AI engage in iterative dialogue to complete a task with the human refining the AI outputs 
- Learning - Human seeks understanding and explanation rather than direct task completion
- Validation - Human uses AI to check or validate their own work

Based on your analysis, identify which of the above is most representative of the user's response. If multiple patterns are present, select the one that is most appears most frequently. If you are unsure or there is not enough context to determine the most representative pattern, return 'None' as your answer. Use 'None' liberally---for only some conversations will this task be possible.

Before providing your final analysis for each response, use the <thinking> tags to break down your reasoning process. Consider the context, the user's words, and how they align with the defined patterns. Provide your output in the following format:

<thinking>
Your 2-3 sentence thinking about your decision. Include specific examples from the conversation and mention any relevant user response patterns observed.
</thinking>
<answer>
Your classification from the list (including only the name of the collaboration pattern) or 'None'
</answer>

Using this information, identify the primary collaboration pattern in the provided conversation. Ensure your classification is the name of a collaboration pattern or 'None'.

Assistant:
\end{spverbatim}

\subsection{Usage by Occupational Skills}
The "shuffle" XML tags in the below conversation shuffle the options provided. The model does not see the shuffle tags when prompted for an assignment.

\begin{spverbatim}
Human: The following is a conversation between Claude, an AI assistant, and a user:

{conversation}

Assistant: I understand.

Human: Please identify which categories best describe the conversation. Consider the provided list of occupational skills. Your job is to identify ALL skills exhibited by the assistant in the following human-AI assistant conversation. You can select multiple skills if appropriate. Select all that apply. Please comma-separate your selections (e.g., 'social perceptiveness', 'science', 'writing', etc.) and provide no additional commentary. If no skills are exhibited by the assistant return 'none'.

<options><shuffle>
reading comprehension
active listening
writing
speaking
mathematics
science
critical thinking
active learning
learning strategies
monitoring
social perceptiveness
coordination
persuasion
negotiation
instructing
service orientation
complex problem solving
operations analysis
technology design
equipment selection
installation
programming
operations monitoring
operation and control
equipment maintenance
troubleshooting
repairing
quality control analysis
judgment and decision making
systems analysis
systems evaluation
time management
management of financial resources
management of material resources
management of personnel resources
none
</shuffle></options>

Assistant:
\end{spverbatim}

\section{Clio Cluster-Based Reconstruction Analyses}
\label{appendix:cluster_experiments}

In our primary analyses above, we used Clio to classify conversations into their best-fit O*NET tasks; these tasks were treated as clusters in Clio to maintain our privacy assurances (importantly, aggregation thresholds and removal of personally identifiable information.). In this section, we replicate our O*NET analysis using a different approach---instead of analyzing individual conversations, we analyze Clio's conversation clusters at different levels of aggregation. Specifically, we assign these clusters to their best-fit O*NET task and examine how the results compare to our conversation-level analysis.

We show that analyzing Clio clusters at higher aggregation levels (where clusters contain many related conversations) produces similar results to analyzing conversations directly for many research questions, such as determining the relative usage of AI across different occupational categories. These results suggest that anonymized, aggregated Clio clusters can serve as a reliable proxy for individual conversation analysis, enabling privacy-preserving study of AI usage patterns while maintaining validity. This analysis also serves as an additional robustness check for the findings presented in the main text of the paper.

\subsection{Method and Sample}

We used Clio to generate clusters that captured \textit{the user's overall request for the assistant}. \cite{clio} provides more information about our clustering process, including full prompts, hyperparameters, and other settings.

We generated clusters from a random sample of approximately 2.8 million Claude.ai Free and Pro conversations between November 28, 2024 and December 18, 2024. Unlike our analyses above, we excluded all conversations that our safety systems deemed abusive or harmful, and we did not exclude conversations that did not appear to relate to an occupational task. We then assigned clusters to their best-fit O*NET task using the same hierarchical process as our conversation-level analysis (for more details about the assignment process, see \Cref{appendix:experimental_details}). As in our primary analysis above, when a task corresponds to multiple occupations, we split the prevalence evenly across those occupations. We applied the same quantitative privacy requirements from our primary analyses. For more information, see \Cref{appendix:experimental_details}.

\subsection{Occupational Categories}

We find strong correlation between the usage of occupational categories determined via direct assignment and via clusters at various aggregation levels. When comparing direct assignment to cluster-based approaches, we observe minimum correlation values of \textbf{0.95} (Pearson), \textbf{0.95} (Spearman), and \textbf{0.82} (Kendall's Tau-b).

\begin{figure}[htpb]
    \centering
    \includegraphics[width=0.95\linewidth]{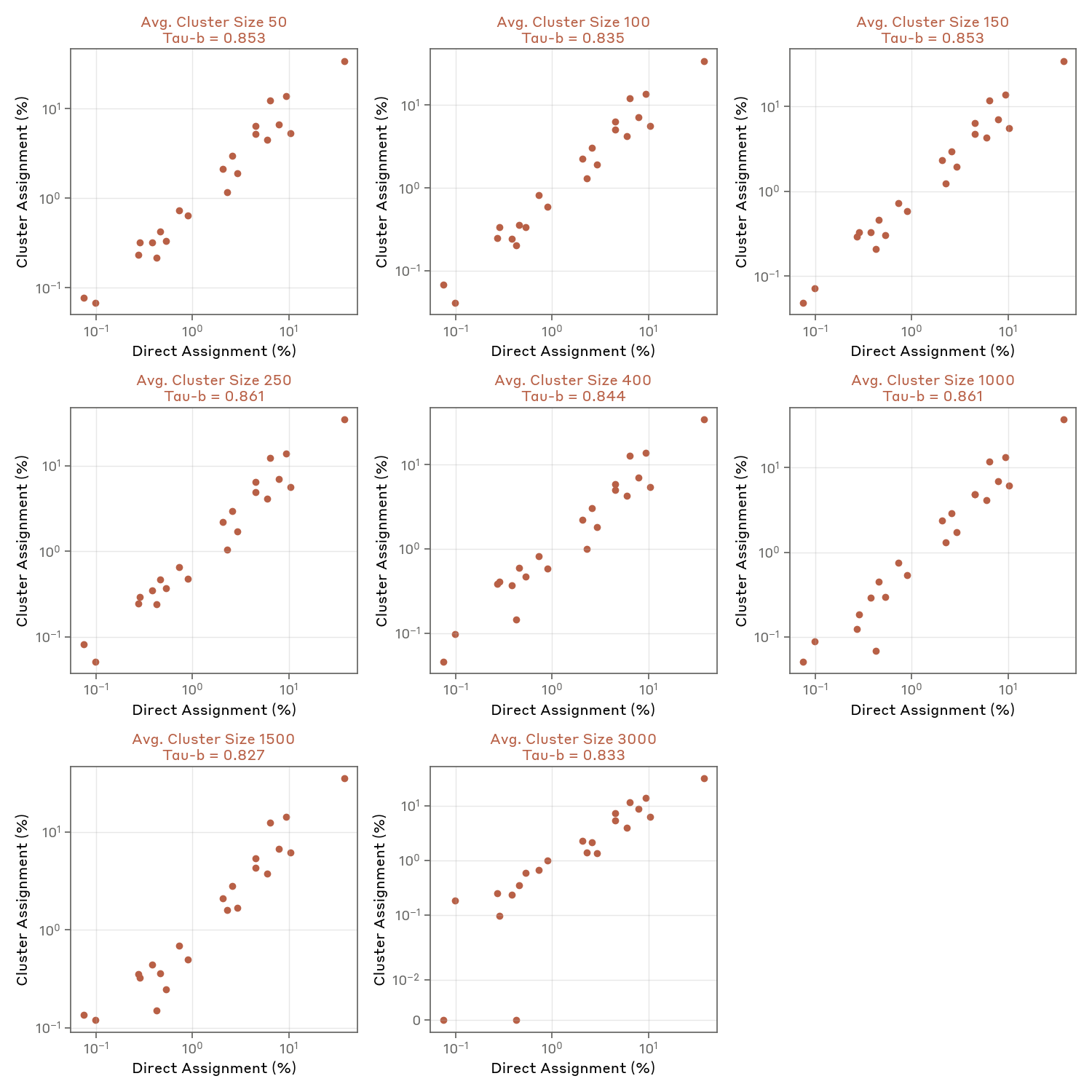}         
    \caption{Comparison between the prevalence of occupational categories when determined via direct assignment compared to clusters at various aggregation levels.}
    \label{fig:occupational_categories_corr}
\end{figure}

\begin{figure}[htpb]
    \centering
    \includegraphics[width=0.95\linewidth]{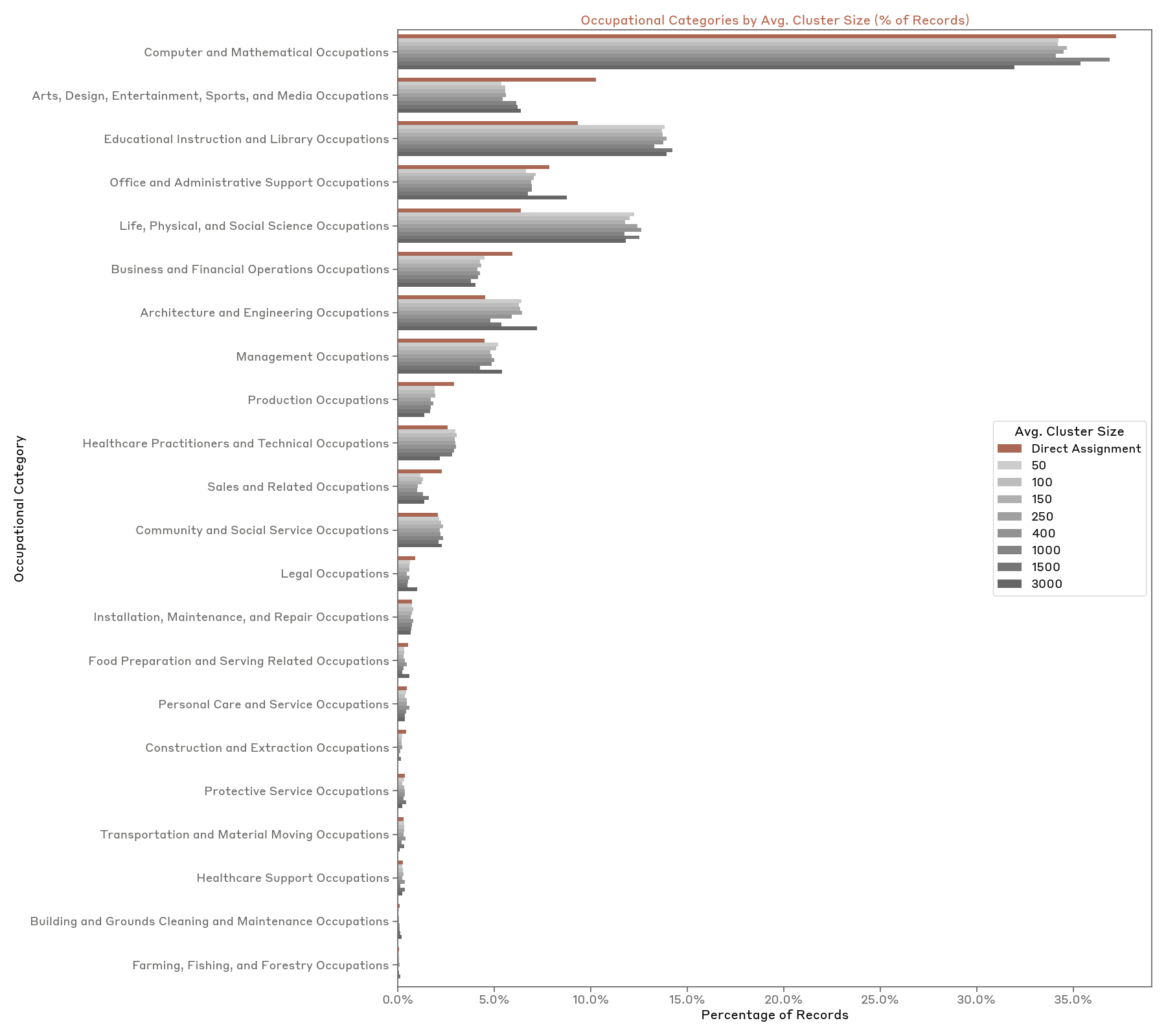}         
    \caption{Prevalence of occupational categories when determined via direct assignment compared to clusters at various aggregation levels.}
    \label{fig:occupational_categories_top}
\end{figure}

\begin{figure}[htpb]
    \centering
    \begin{subfigure}[b]{0.45\linewidth}
        \centering
        \includegraphics[width=\linewidth]{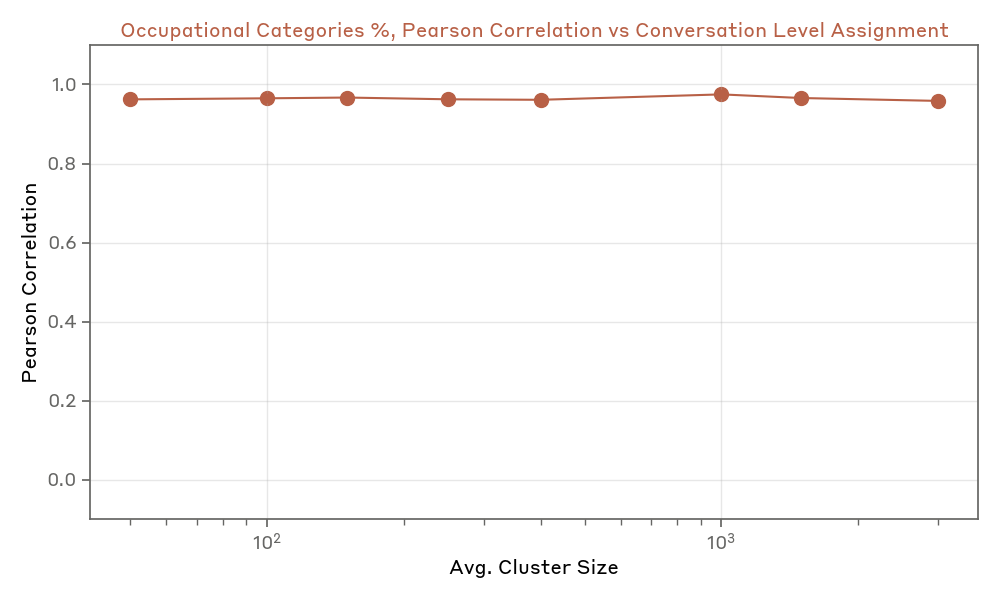}
    \end{subfigure}
    \begin{subfigure}[b]{0.45\linewidth}
        \centering
        \includegraphics[width=\linewidth]{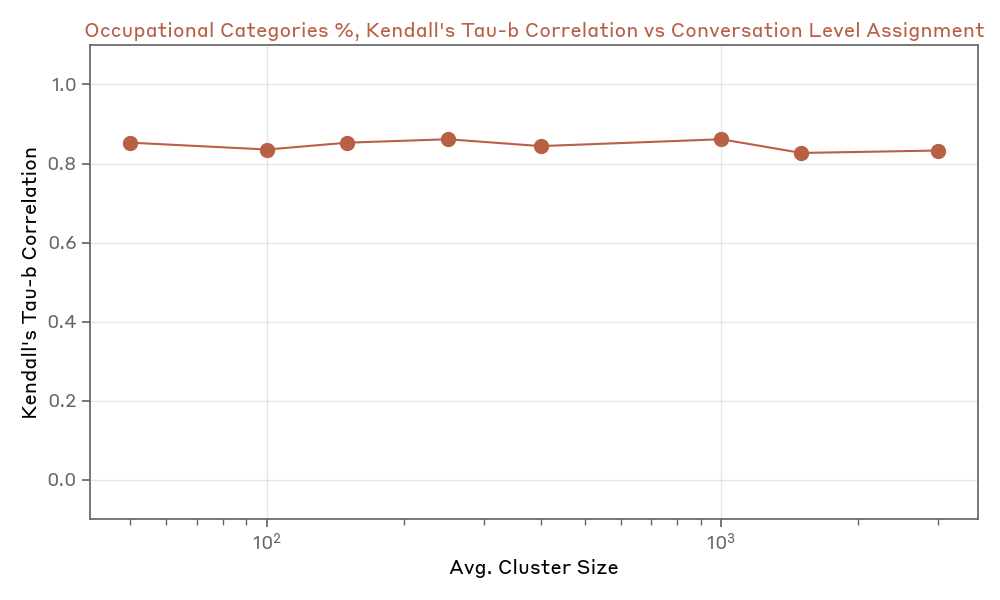}
    \end{subfigure}
    \vspace{1em}
    \begin{subfigure}[b]{0.45\linewidth}
        \centering
        \includegraphics[width=\linewidth]{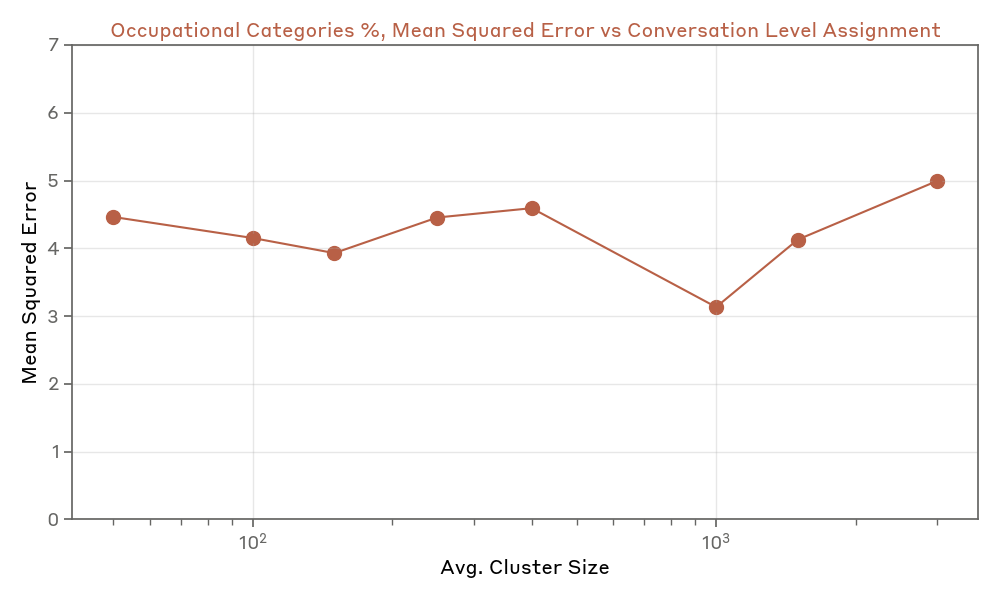}
    \end{subfigure}
    \begin{subfigure}[b]{0.45\linewidth}
        \centering
        \includegraphics[width=\linewidth]{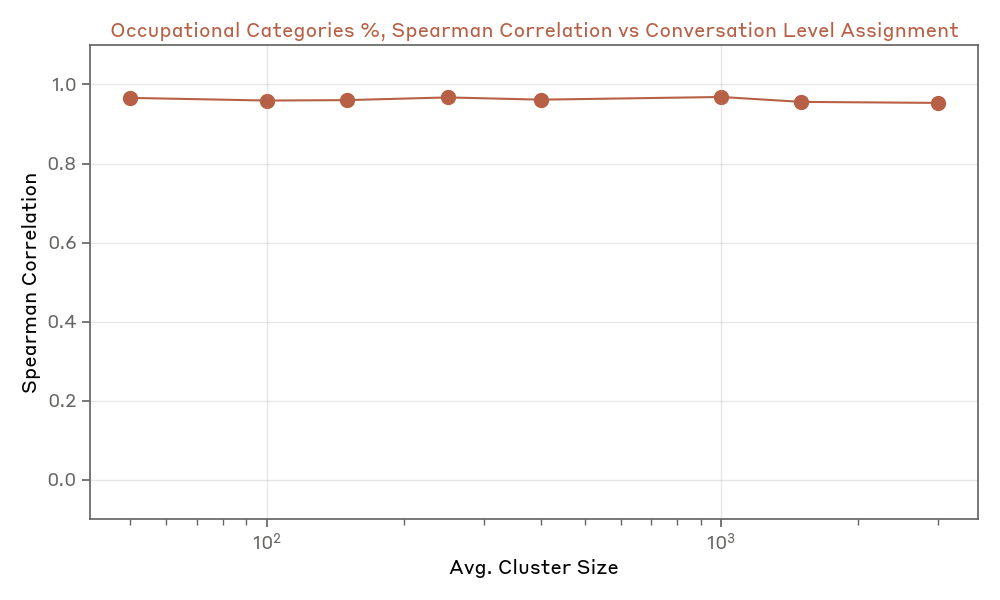}
    \end{subfigure}
    \caption{Comparison of occupational category distributions between direct assignment and cluster-based approaches at various aggregation levels, evaluated using different correlation metrics (Pearson, Kendall, Spearman) and Mean Square Error (MSE). These metrics provide complementary views of how well the cluster-based categorization aligns with analysis on conversations directly.}
    \label{fig:occupational_categories_metrics}
\end{figure}

We visualize these relationships in several ways. \Cref{fig:occupational_categories_corr} directly compares the prevalence of occupational categories between direct assignment and cluster-based approaches at different aggregation levels. \Cref{fig:occupational_categories_top} shows how category prevalence varies across all aggregation levels, while \Cref{fig:occupational_categories_metrics} presents the correlation metrics alongside mean squared error measurements.

\subsection{Occupation Reconstruction}

We find reasonable correlation between occupational usage patterns whether measured through direct conversation assignment or through cluster-based assignment at different aggregation levels. Specifically, when comparing direct assignment to cluster-based approaches, we observe minimum correlation values of \textbf{0.70} (Pearson), \textbf{0.58} (Spearman), and \textbf{0.46} (Kendall's Tau-b). These correlations are calculated only for occupations that have associated tasks identified by at least one method.

\begin{figure}[htpb]
    \centering
    \includegraphics[width=0.95\linewidth]{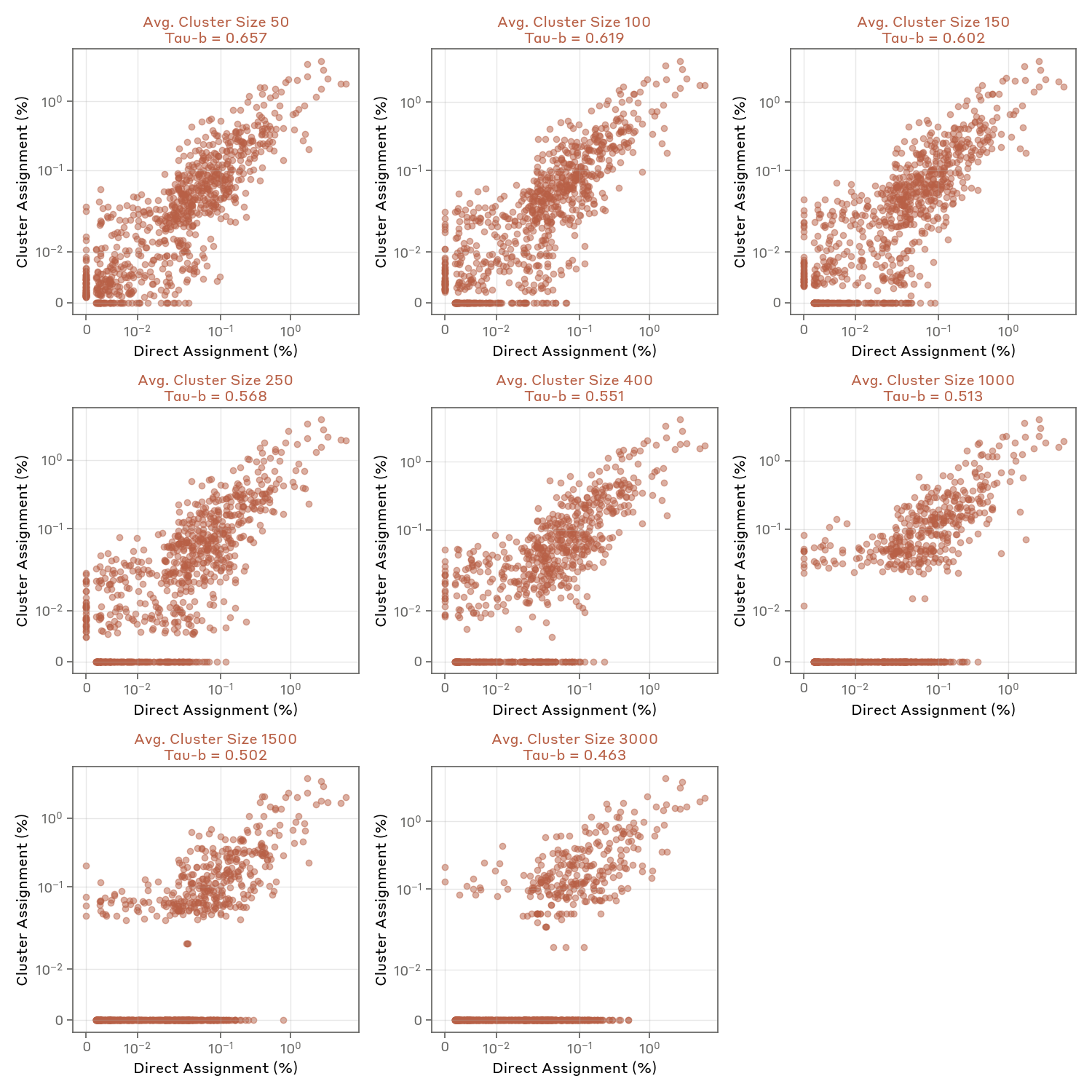}         
    \caption{Comparison between the prevalence of occupations when determined via direct assignment compared to clusters at various aggregation levels.}
    \label{fig:occupation_corr}
\end{figure}

\begin{figure}[htpb]
    \centering
    \includegraphics[width=0.95\linewidth]{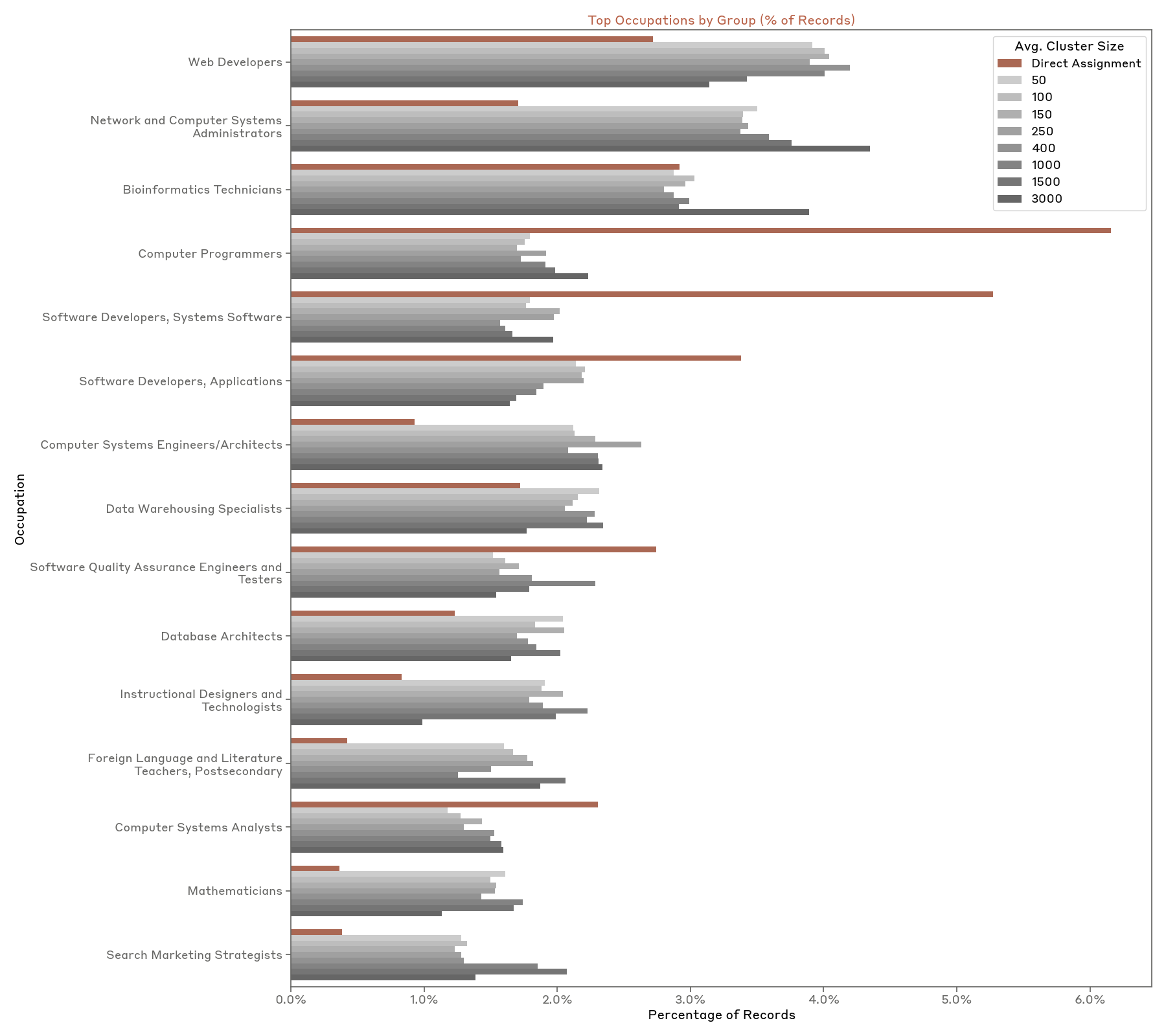}         
    \caption{Prevalence of top occupations when determined via direct assignment compared to clusters at various aggregation levels.}
    \label{fig:top_occupations}
\end{figure}

\begin{figure}[htpb]
    \centering
    \begin{subfigure}[b]{0.45\linewidth}
        \centering
        \includegraphics[width=\linewidth]{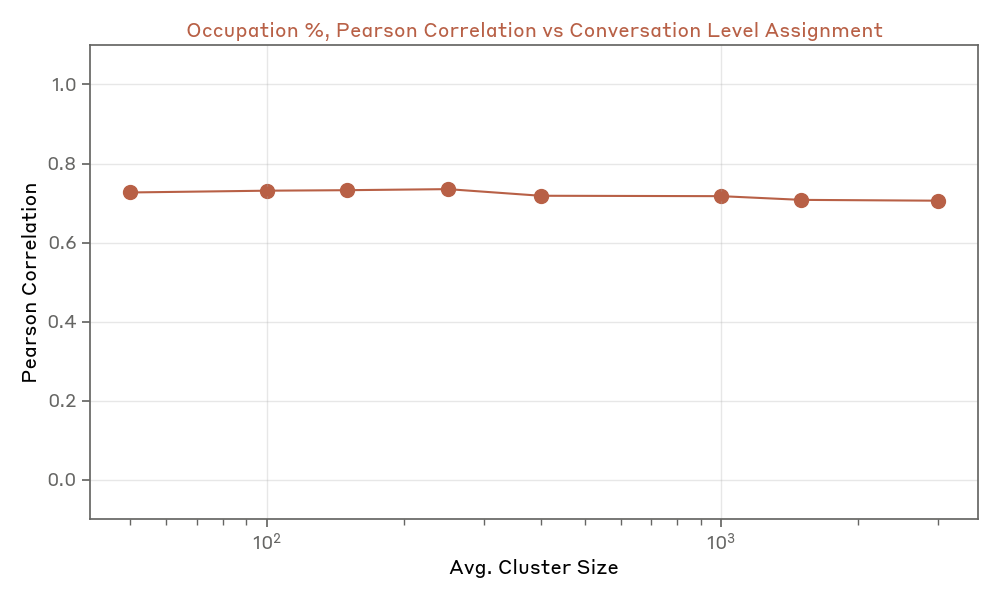}
    \end{subfigure}
    \begin{subfigure}[b]{0.45\linewidth}
        \centering
        \includegraphics[width=\linewidth]{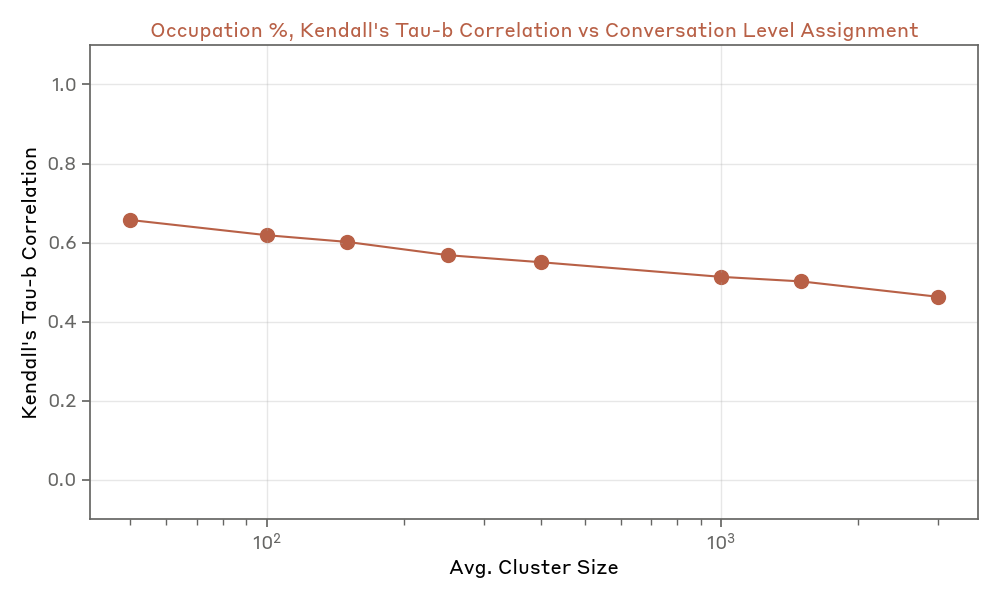}
    \end{subfigure}
    \vspace{1em}
    \begin{subfigure}[b]{0.45\linewidth}
        \centering
        \includegraphics[width=\linewidth]{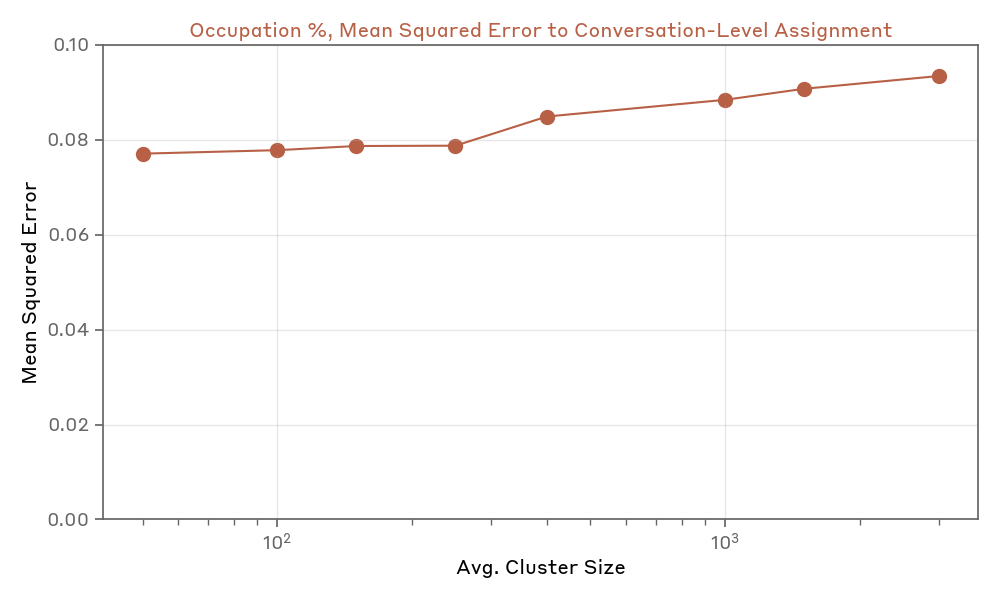}
    \end{subfigure}
    \begin{subfigure}[b]{0.45\linewidth}
        \centering
        \includegraphics[width=\linewidth]{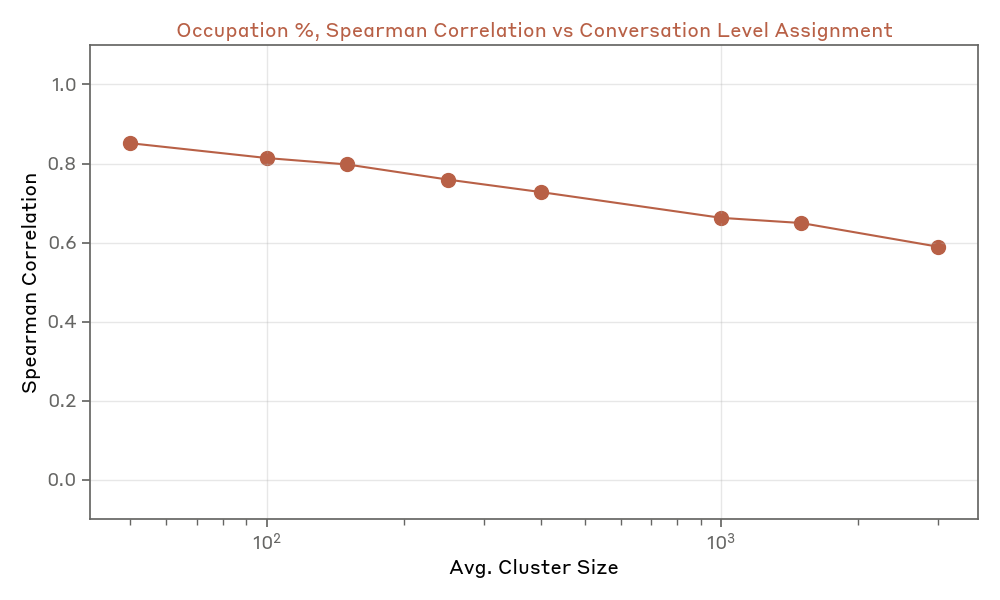}
    \end{subfigure}
    \caption{Comparison of occupation distributions between direct assignment and cluster-based approaches at various aggregation levels, evaluated using different correlation metrics (Pearson, Kendall, Spearman) and Mean Square Error (MSE). These metrics provide complementary views of how well the cluster-based categorization aligns with analysis on conversations directly.}
    \label{fig:occupation_metrics}
\end{figure}

\begin{figure}[htpb]
    \centering
    \begin{subfigure}[b]{0.8\linewidth}
        \centering
        \includegraphics[width=\linewidth]{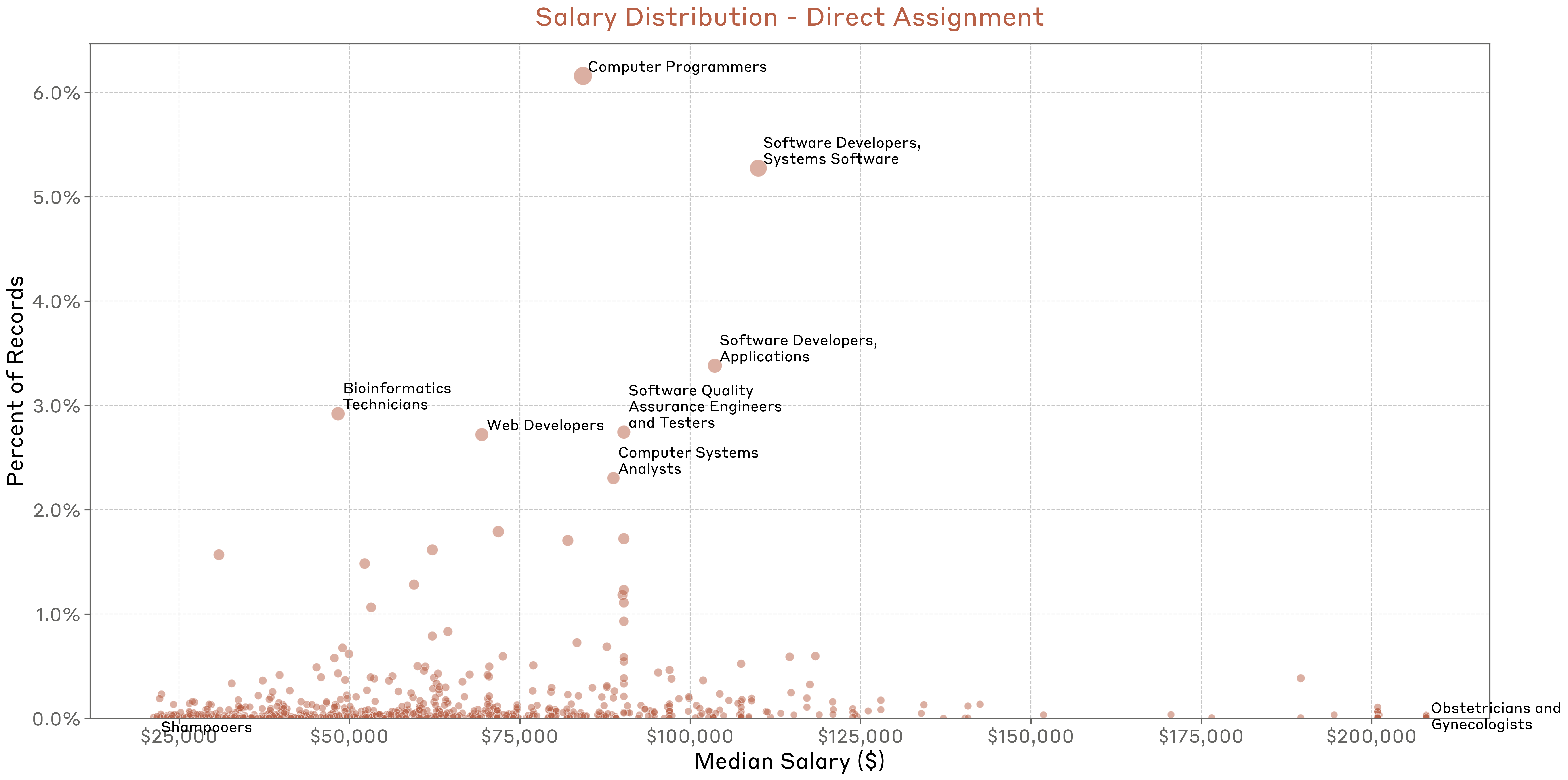}
    \end{subfigure}
    \begin{subfigure}[b]{0.8\linewidth}
        \centering
        \includegraphics[width=\linewidth]{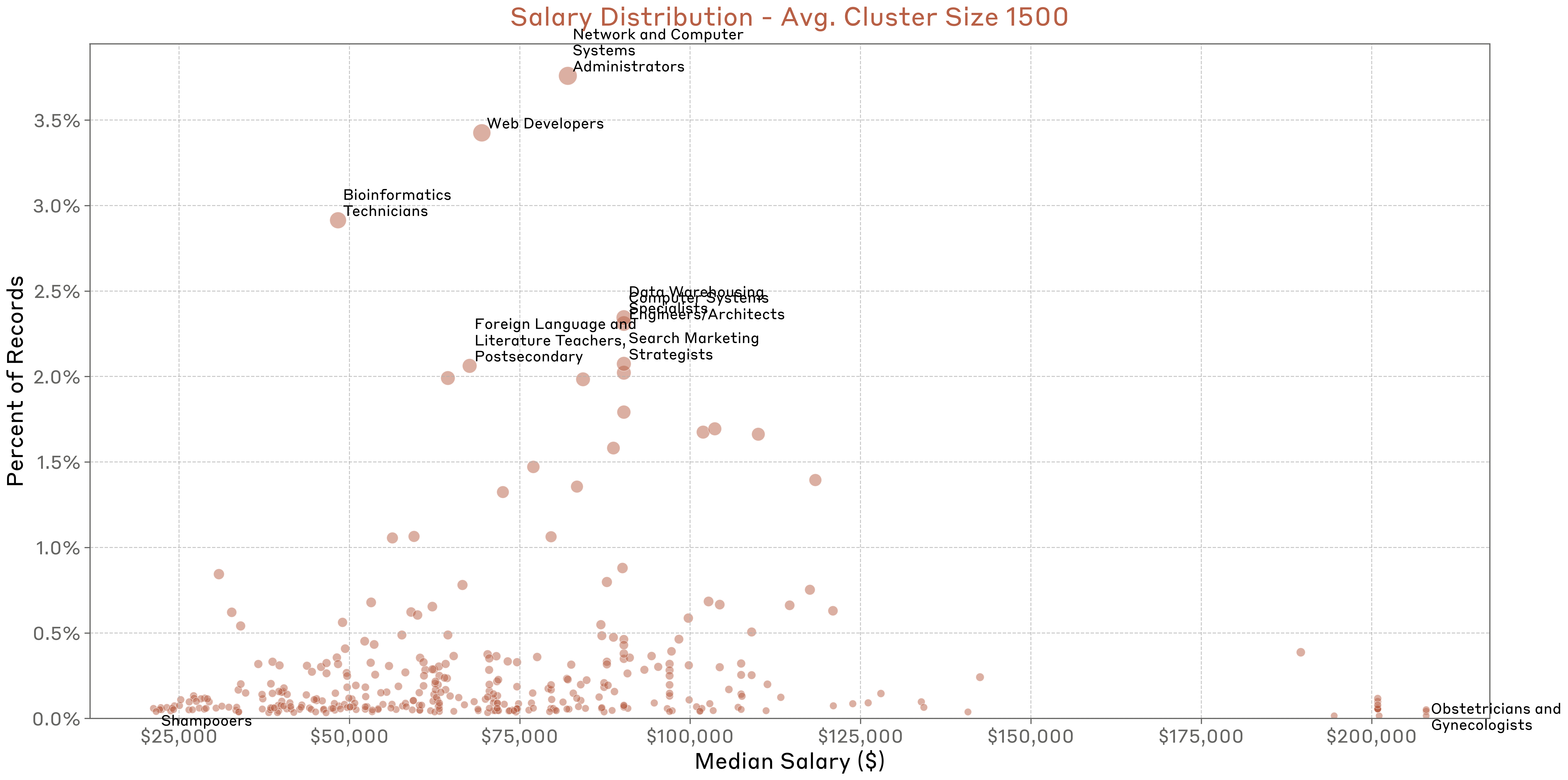}
    \end{subfigure}
    \caption{Comparison of median salary by task usage between direct assignment and clusters of size $\sim\!1,500$. In both cases, task usage is highest for occupations with a median salary between \$50,000 and \$125,000.}
    \label{fig:salary_aggregation_distribution}
\end{figure}

We visualize these relationships in several ways. \Cref{fig:occupation_corr} directly compares the prevalence of occupations between direct assignment and cluster-based approaches at different aggregation levels. \Cref{fig:top_occupations} shows how occupation prevalence varies across all aggregation levels, while \Cref{fig:occupation_metrics} presents the correlation metrics alongside mean squared error measurements. \Cref{fig:salary_aggregation_distribution} shows the difference in usage by median salary between direct assignment and cluster assignment.

\begin{figure}[htpb]
    \centering
    \includegraphics[width=0.95\linewidth]{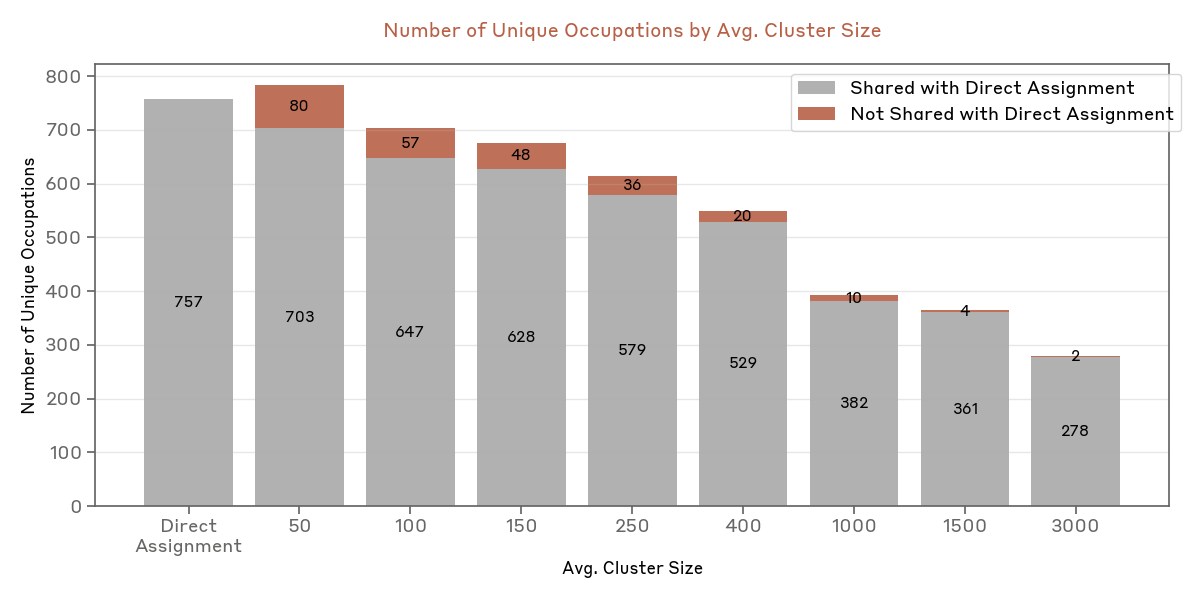}         
    \caption{Number of occupations recovered at each aggregation level compared to direct assignment.}
    \label{fig:occupation_recovery}
\end{figure}

Given the significant overlap between many tasks and occupations in the O*NET database, we expect that the two methods would identify somewhat different sets of tasks and their associated occupations. \Cref{fig:occupation_recovery} illustrates this pattern, showing which occupations were identified by each method and where these sets overlap.

\subsection{Task Reconstruction}

We find relatively weak correlation between task usage patterns when comparing direct assignment to cluster-based approaches. When comparing these methods, we observe minimum correlation values of \textbf{0.47} (Pearson), \textbf{0.22} (Spearman), and \textbf{0.18} (Kendall's Tau-b). These correlations are calculated only for tasks identified by at least one method.

\begin{figure}[htpb]
    \centering
    \includegraphics[width=0.95\linewidth]{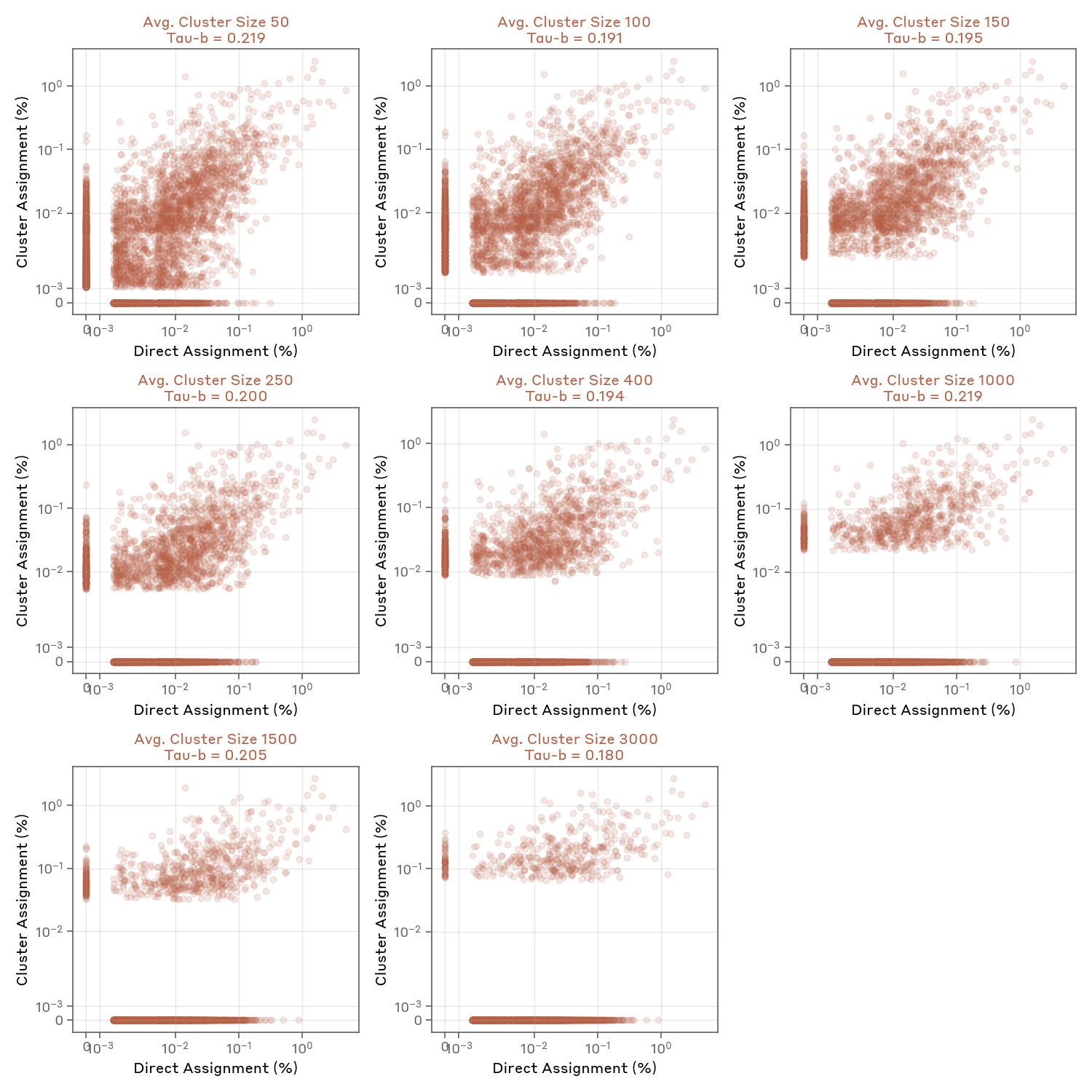}         
    \caption{Comparison between the prevalence of tasks when determined via direct assignment compared to clusters at various aggregation levels.}
    \label{fig:task_corr}
\end{figure}

\begin{figure}[htpb]
    \centering
    \includegraphics[width=0.95\linewidth]{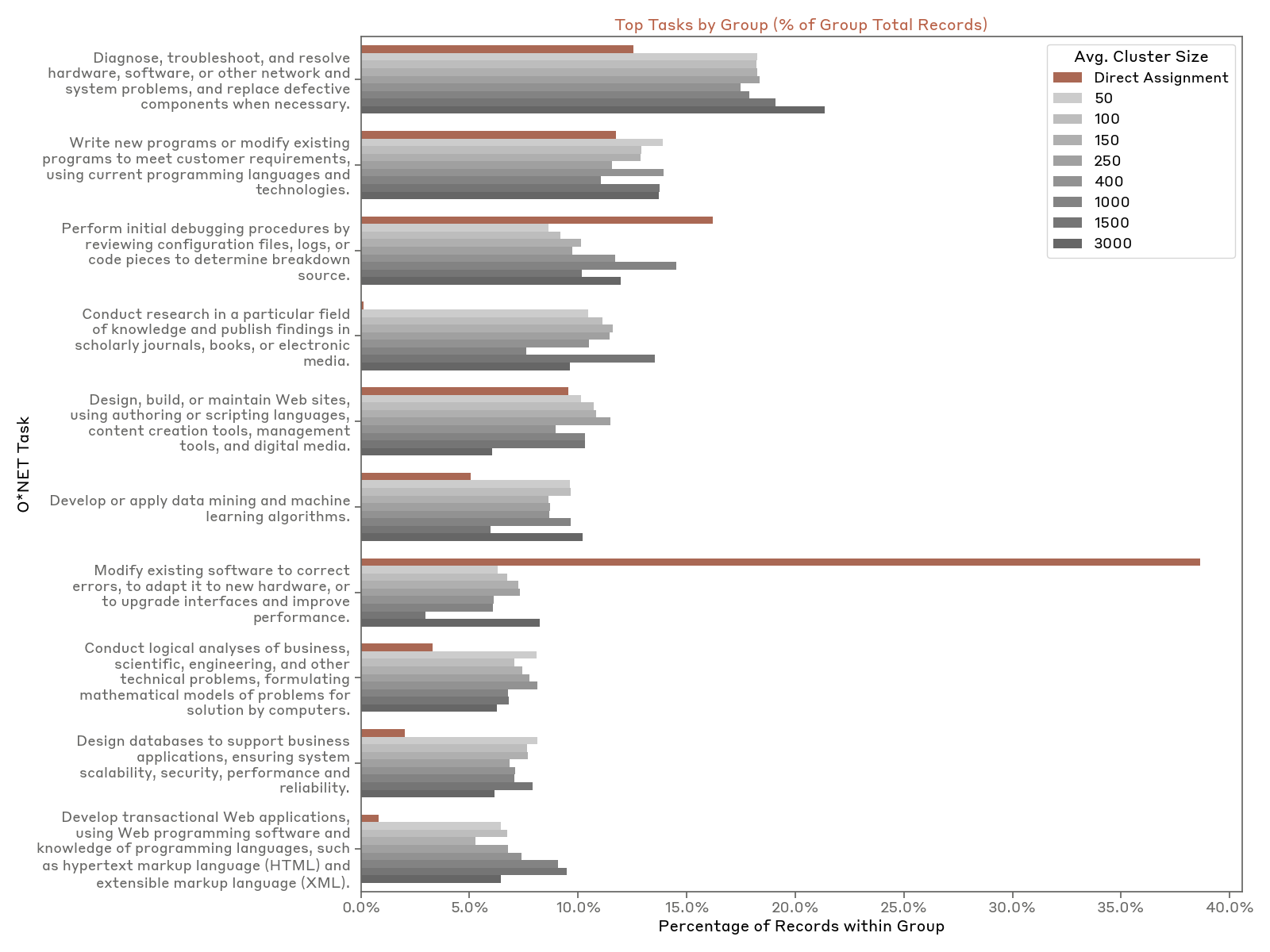}         
    \caption{Prevalence of top tasks when determined via direct assignment compared to clusters at various aggregation levels.}
    \label{fig:reconstruction_top_tasks}
\end{figure}

\begin{figure}[htpb]
    \centering
    \begin{subfigure}[b]{0.45\linewidth}
        \centering
        \includegraphics[width=\linewidth]{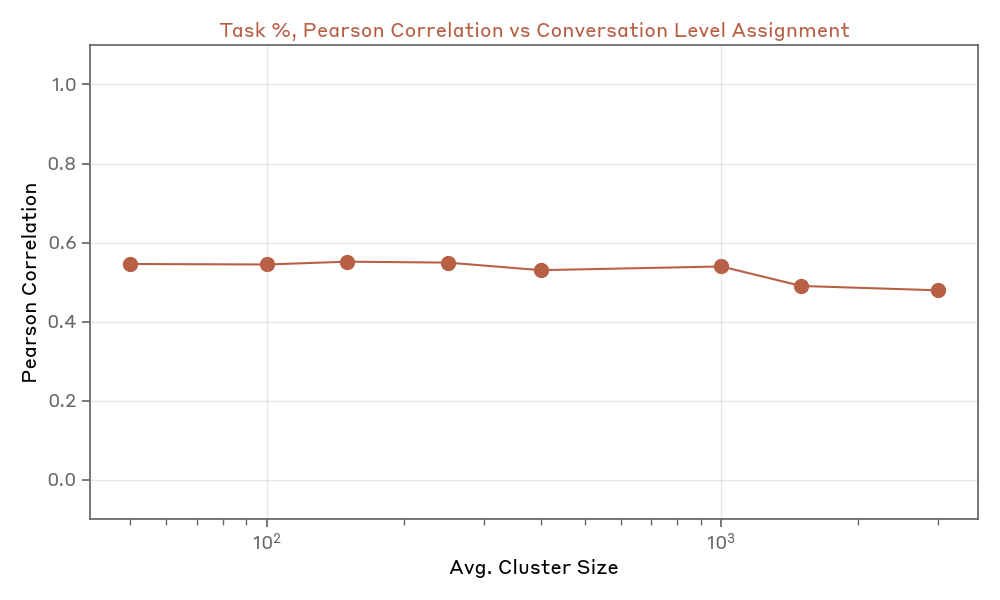}
    \end{subfigure}
    \begin{subfigure}[b]{0.45\linewidth}
        \centering
        \includegraphics[width=\linewidth]{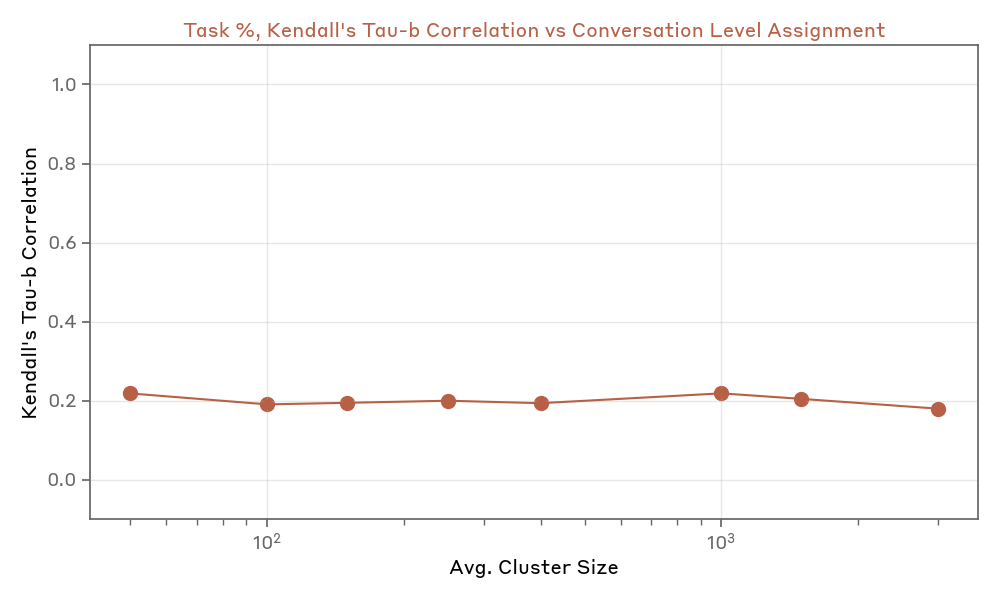}
    \end{subfigure}
    \vspace{1em}
    \begin{subfigure}[b]{0.45\linewidth}
        \centering
        \includegraphics[width=\linewidth]{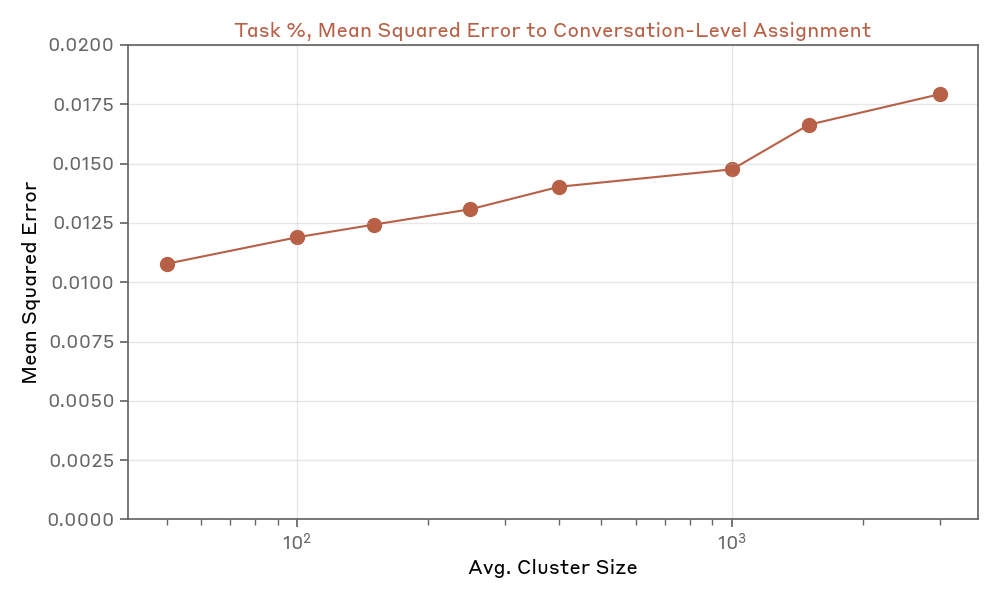}
    \end{subfigure}
    \begin{subfigure}[b]{0.45\linewidth}
        \centering
        \includegraphics[width=\linewidth]{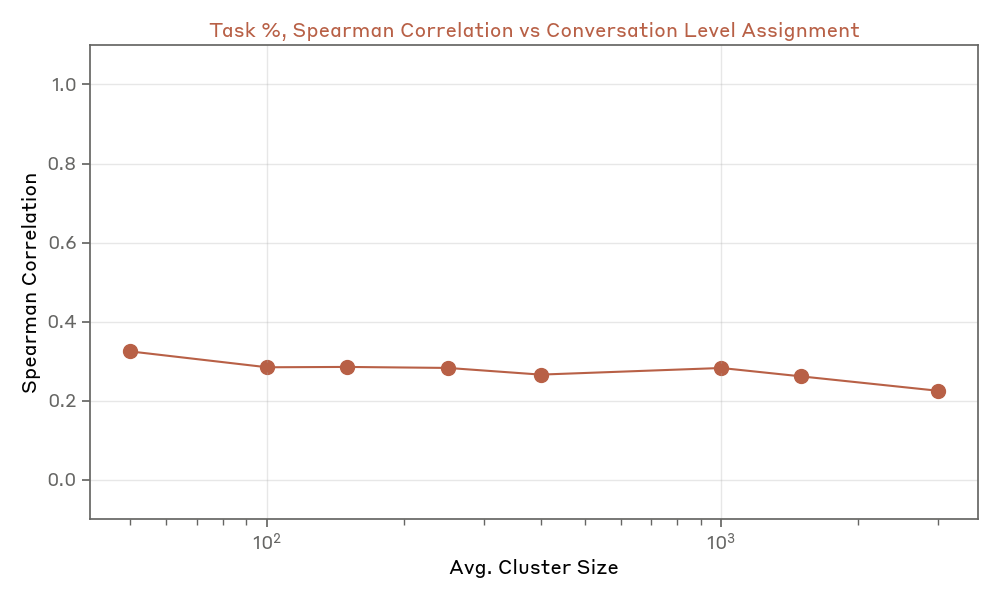}
    \end{subfigure}
    \caption{Comparison of task distributions between direct assignment and cluster-based approaches at various aggregation levels, evaluated using different correlation metrics (Pearson, Kendall, Spearman) and Mean Square Error (MSE). These metrics provide complementary views of how well the cluster-based categorization aligns with analysis on conversations directly.}
    \label{fig:task_metrics}
\end{figure}

We visualize these relationships in several ways. \Cref{fig:task_corr} directly compares the prevalence of tasks between direct assignment and cluster-based approaches at different aggregation levels. \Cref{fig:reconstruction_top_tasks} shows how task prevalence varies across all aggregation levels, while \Cref{fig:task_metrics} presents the correlation metrics alongside mean squared error measurements.

The relatively weak correlation at the task level can be attributed to two key factors. First, there is intrinsic noise in the O*NET task dataset itself, with many tasks being highly similar or overlapping. For example, tasks like "Diagnose, troubleshoot, and resolve hardware, software, or other network system problems, and replace defective components when necessary" and "Perform initial debugging procedures by reviewing configuration files, logs, or code pieces to determine breakdown source" are functionally very similar but listed as distinct tasks. Second, there exists a many-to-many relationship between conversations and tasks (a single conversation might involve multiple tasks) and between tasks and occupations (a single task might map to multiple occupations). This means that uncertainty expands at the task level but naturally narrows as we aggregate upward to occupational categories; therefore, we see stronger correlations at higher levels of aggregation. This pattern suggests that while cluster-based analysis may not perfectly reconstruct task-level assignments, it remains reliable for analyzing broader occupational trends.

\begin{figure}[htpb]
    \centering
    \includegraphics[width=0.95\linewidth]{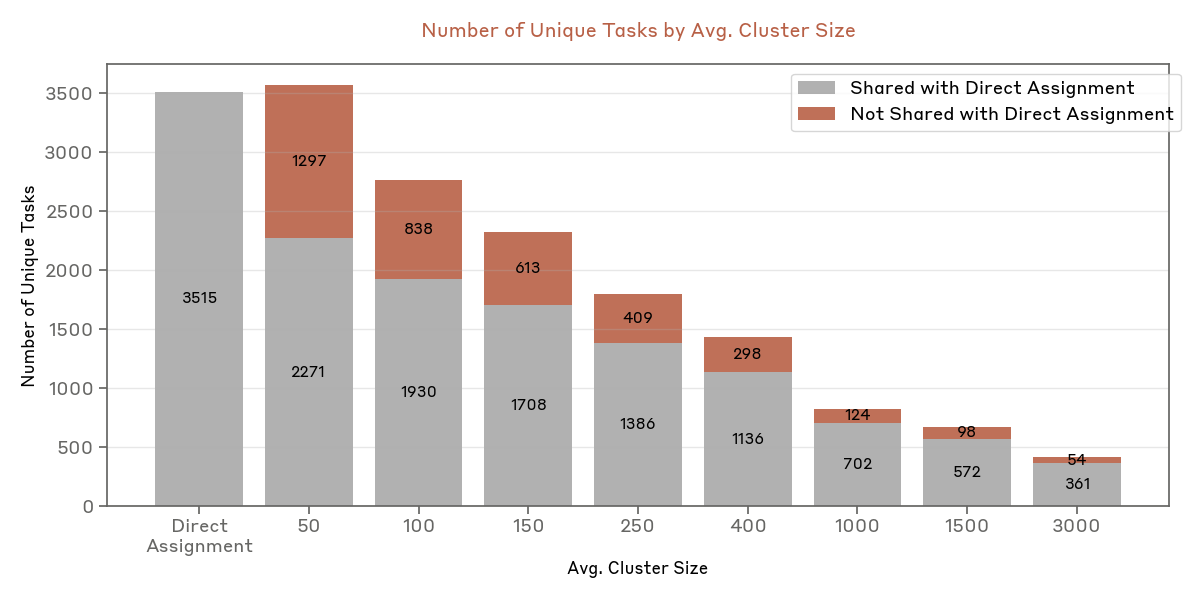}         
    \caption{Number of occupations recovered at each aggregation level compared to direct assignment. Notably, only a small fraction (<20\%) of the $\sim\!20$K O*NET tasks are recovered.}
    \label{fig:task_recovery}
\end{figure}

\begin{figure}[htpb]
    \centering
    \includegraphics[width=0.95\linewidth]{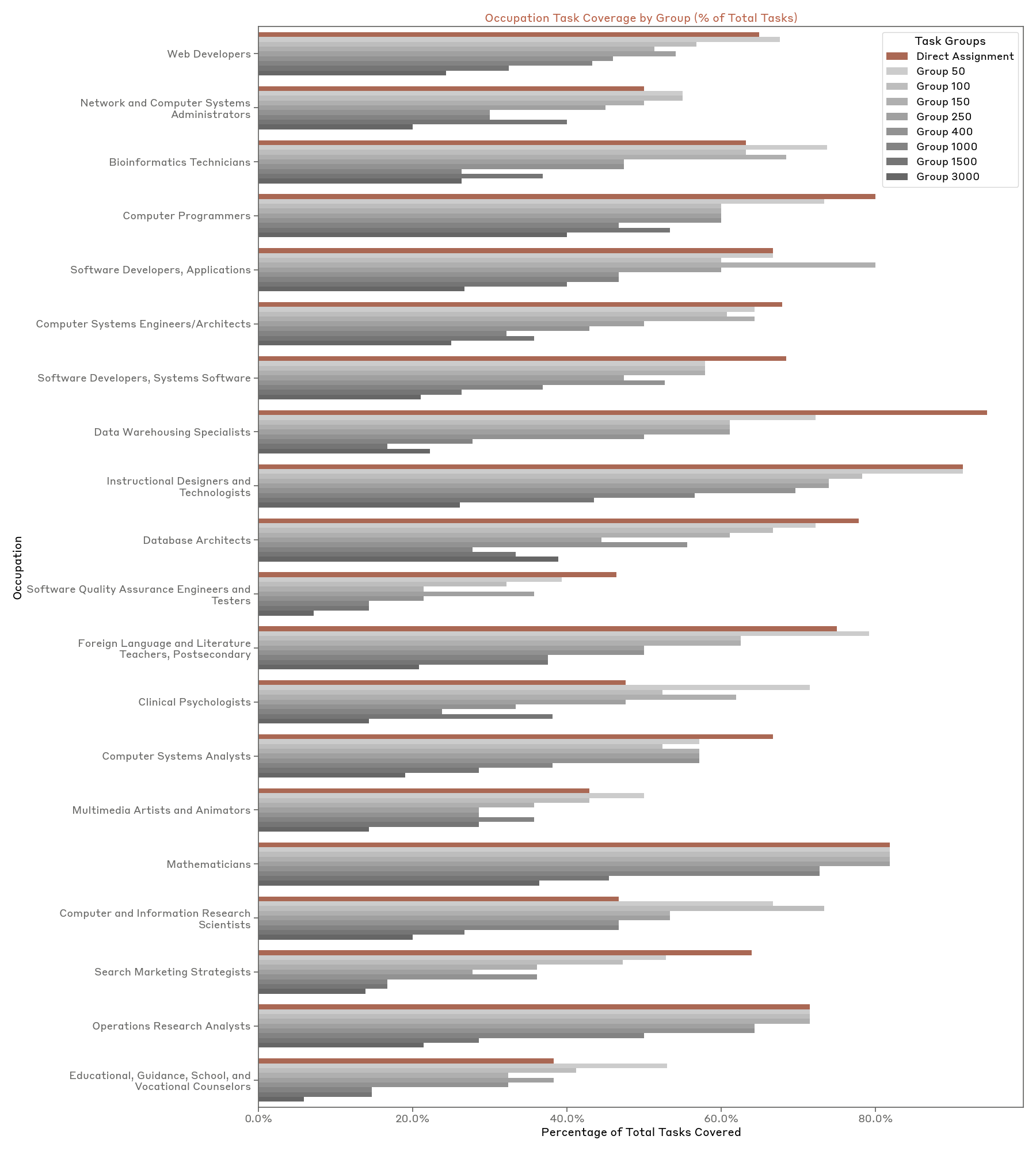}         
    \caption{Number of tasks assigned to top occupations at various aggregation levels. As expected, higher aggregation levels have fewer average tasks per occupation.}
    \label{fig:occupation_task_coverage_by_frequency}
\end{figure}

\begin{figure}[htpb]
    \centering
    \includegraphics[width=0.95\linewidth]{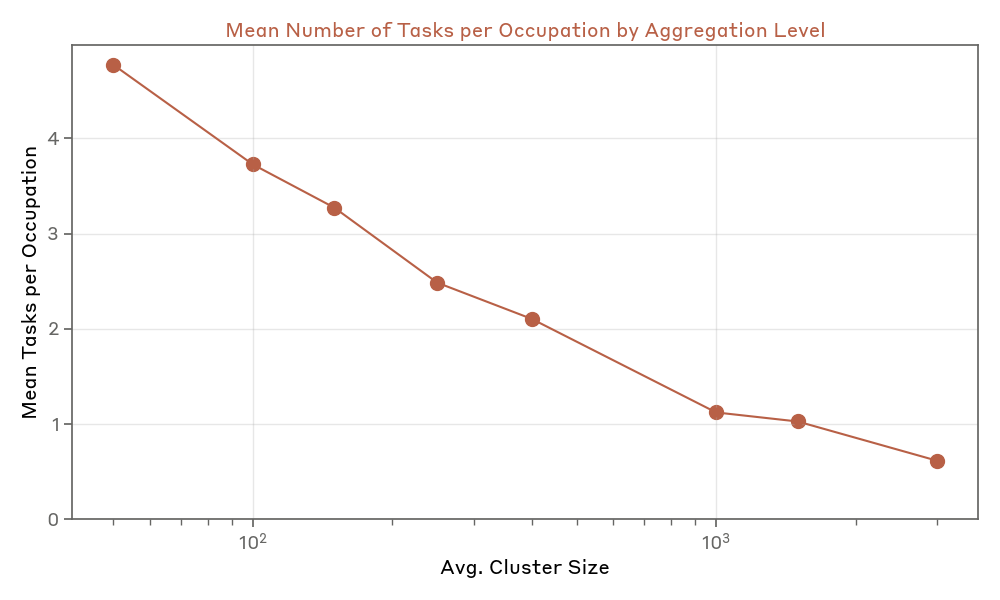}         
    \caption{Mean number of tasks assigned to each occupation at various aggregation levels. Direct assignment assigns an average of 4.8 tasks per occupation.}
    \label{fig:mean_tasks_per_occupation}
\end{figure}

Given the granular nature of individual tasks in the O*NET database and significant overlap between similar tasks, we expect considerable variation in which specific tasks are identified by each method. \Cref{fig:task_recovery} illustrates this pattern, showing which tasks were identified by each method and where these sets overlap. \Cref{fig:occupation_task_coverage_by_frequency} and \Cref{fig:mean_tasks_per_occupation} show the average number of tasks assigned to tasks at various aggregation levels.

\end{document}